# Source Coding With Encoder Side Information

Emin Martinian, Gregory Wornell, Ram Zamir

**Abstract**

We introduce the idea of distortion side information, which does not directly depend on the source but instead affects the distortion measure. We show that such distortion side information is not only useful at the encoder, but that under certain conditions, knowing it at only the encoder is as good as knowing it at both encoder and decoder, and knowing it at only the decoder is useless. Thus distortion side information is a natural complement to the signal side information studied by Wyner and Ziv, which depends on the source but does not involve the distortion measure. Furthermore, when both types of side information are present, we characterize the penalty for deviating from the configuration of encoder-only distortion side information and decoder-only signal side information, which in many cases is as good as full side information knowledge.

## I. INTRODUCTION

In many large systems such as sensor networks, communication networks, distributed control, and biological systems different parts of the system may each have limited, noisy, or incomplete information but must somehow cooperate. Key issues in such scenarios include the penalty incurred due to the lack of shared information, possible approaches for combining information from different sources, and the more general question of how different kinds of information can be partitioned based on the role of each system component.

One example of this scenario illustrated in Fig. 1(a), is when an observer records a signal **x** to be conveyed to a receiver who also has some additional *signal side information* **w**, which is correlated with **x**. As first introduced by Wyner and Ziv [4], [5] and extended by other researchers [6], [7], [8], in many cases the observer and receiver can obtain the full benefit of the signal side information even if it is known only by the receiver.

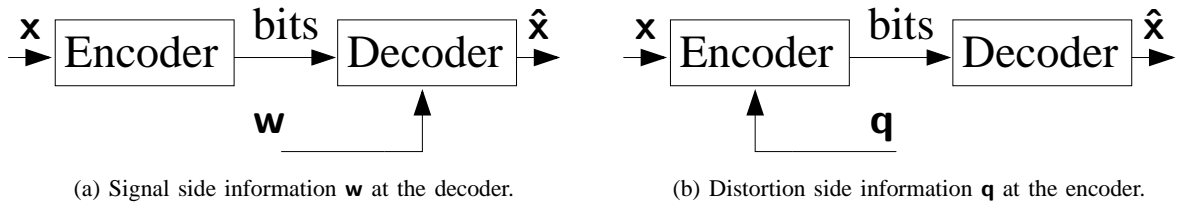

(a) Signal side information **w** at the decoder.  (b) Distortion side information **q** at the encoder.

Fig. 1. Compressing a source **x** with signal side information or distortion side information.

In this paper, we introduce a different scenario illustrated in Fig. 1(b) where instead the observer has some *distortion side information* **q** describing which components of the data are more important than others, but the receiver may not have access to **q**. Specifically, let us model the differing importance of different signal components by measuring the distortion between the $i$th source sample $x[i]$ and its quantized value $\hat{x}[i]$ by a distortion function that depends on the side information $q[i]$:

$$d(x[i], \hat{x}[i]; q[i]). \tag{1}$$

In principle, one could treat the source-side information pair $(\mathbf{q}, \mathbf{x})$ as an "effective composite source", and apply conventional techniques to quantize it. Such an approach, however, ignores the different roles **q** and **x** play in the distortion. And as often happens in lossy compression, a good understanding of the distortion measure may lead to a more efficient system. Moreover, an interesting observation made in this paper is that in some important cases the encoder can obtain the full benefit of the distortion side information even if it is not known at the receiver. Hence, distortion side information at the encoder is a natural complement of the Wyner-Ziv setting.

Sensor observations are one class of signals where the idea of distortion side information may be useful. For example, a sensor may have side information corresponding to reliability estimates for measured data (which may or may not be available at the receiver). This may occur if the sensor can calibrate its accuracy to changing conditions (*e.g.*, the amount of light, background noise, or other interference present), if the sensor averages data for a variety of measurements (*e.g.*, combining results from a number of sub-sensors) or if some external signal indicates important events (*e.g.*, an accelerometer indicating movement).

Alternatively, certain components of the signal may be more or less sensitive to distortion due to masking effects or context [9]. For example errors in audio samples following a loud sound, or errors in pixels spatially or temporally near bright spots are perceptually less relevant. Similarly, accurately preserving certain edges or textures in an image or human voices in audio may be more important than preserving background patterns/sounds. Masking, sensitivity to context, etc., is usually a complicated function of the entire signal. Yet often there is no need to explicitly convey information about this function to the encoder. Hence, from the point of view of quantizing a given sample, it is reasonable to model such effects as side information which is roughly statistically independent of that sample.

Clearly in performing data compression with distortion side information, the encoder should more heavily weight matching the more important data. The importance of exploiting the different sensitivities of the human perceptual system is widely recognized by engineers involved in the construction and evaluation of practical compression algorithms *when distortion side information is available at both observer and receiver*. In contrast, the value and use of distortion side information known only at either the encoder or decoder but not both has received relatively little attention in the information theory and quantizer design community. The rate-distortion function with decoder-only side information, relative to side information dependent distortion measures (as an extension of the Wyner-Ziv setting [4]), is given in [7]. A high resolution approximation for this rate-distortion function for locally quadratic weighted distortion measures is given in [10].

We are not aware of an information-theoretic treatment of encoder-only side information with such distortion measures. In fact, the mistaken notion that encoder-only side information is never useful is common folklore. This may be due to a misunderstanding of Berger's result that side information *that does not affect the distortion measure* is never useful when known only at the encoder [11], [6].

In this paper, we begin by studying the rate-distortion trade-off when side information about the distortion sensitivity is available. We show that such distortion side information can provide an arbitrarily large advantage (relative to no side information) even when the distortion side information is known only at the encoder. Furthermore, we show that just as knowledge of signal side information is often only required at the decoder, knowledge of distortion side information is often only required at the encoder. Finally, we show that these results continue to hold even when both distortion side information **q** and signal side information **w** are considered. Specifically, we demonstrate that a system where only the encoder knows **q** and only the decoder knows **w** is asymptotically as good as a system with all side information known everywhere. We also derive the penalty for deviating from this side information configuration (*e.g.*, providing **q** to the decoder instead of the encoder).

We first illustrate how distortion side information can be used even when known only by the observer with some motivating examples in Section II. Next, in Section III, we precisely define a problem model and state the relevant rate-distortion trade-offs. In Section IV we consider scenarios where encoder-only knowledge of the distortion side information is optimal. Specifically, we show that sources that are uniformly distributed over a group with a difference distortion measure as well as arbitrary sources with "erasure distortion" have the property that encoder-only distortion side information is just as good as full distortion side information. In Section V, we study more general source and distortion models in the limit of high-resolution. Specifically, we show that in high-resolution, knowing distortion side information at the encoder and signal side information at the decoder is both necessary and sufficient to achieve the performance of a fully informed system. In Section VI we consider scaled quadratic distortions in the non-asymptotic (general resolution) regime and derive bounds on the loss with encoder-only distortion side information and decoder-only signal side information versus full side information. These bounds also show how quickly the high resolution regime is approached. Finally, we close with a discussion in Section VII followed by some concluding remarks in Section VIII. Throughout the paper, most proofs and lengthy derivations are deferred to the appendix.

## II. MOTIVATING EXAMPLES

### A. Discrete Uniform Source

Consider the case where the source $x[i]$ corresponds to $n$ samples each uniformly and independently drawn from the finite alphabet $\mathcal{X}$ with cardinality $|\mathcal{X}| \geq n$. Let $q[i]$ correspond to $n$ binary variables indicating which source samples are relevant. Specifically, let the distortion measure be of the form $d(x, \hat{x}; q) = 0$ if and only if either $q = 0$ or $x = \hat{x}$. Finally, let the sequence $q[i]$ be statistically independent of the source with $q[i]$ drawn uniformly from the $n$ choose $k$ subsets with exactly $k$ ones.[1]

If the side information were unavailable or ignored, then losslessly communicating the source would require exactly $n \cdot \log |\mathcal{X}|$ bits. A better (though still sub-optimal) approach when encoder side information is available would be for the encoder to first tell the decoder which samples are relevant and then send only those samples. Using Stirling's approximation, this would require about $n \cdot H_b(k/n)$ bits (where $H_b(\cdot)$ denotes the binary entropy function) to describe which samples are relevant plus $k \cdot \log |\mathcal{X}|$ bits to describe the relevant source samples. Note that if the side information were also known at the decoder, then the overhead required in telling the decoder which samples are relevant could be avoided and the total rate required would only be $k \cdot \log |\mathcal{X}|$. This overhead can in fact be avoided even without decoder side information.

---

[1] If the distortion side information is a Bernoulli($k/n$) sequence, then there will be about $k$ ones with high probability. We focus on the case with exactly $k$ ones for simplicity.

To see this, we view the source samples $x[0]$, $x[1]$, ..., $x[n-1]$, as a codeword of an $(n,k)$ Reed-Solomon (RS) code (or more generally any MDS[2] code) with $q[i] = 0$ indicating an erasure at sample $i$. We use the RS *decoding* algorithm to "correct" the erasures and determine the $k$ corresponding information symbols, which are sent to the receiver. To reconstruct the signal, the receiver *encodes* the $k$ information symbols using the encoder for the $(n,k)$ RS code to produce the reconstruction $\hat{x}[0]$, $\hat{x}[1]$, ..., $\hat{x}[n-1]$. Only symbols with $q[i] = 0$ could have changed, hence $\hat{x}[i] = x[i]$ whenever $q[i] = 1$ and the relevant samples are losslessly communicated using only $k \cdot \log|\mathcal{X}|$ bits.

As illustrated in Fig. 2, RS decoding can be viewed as curve-fitting and RS encoding can be viewed as interpolation. Hence this source coding approach can be viewed as fitting a curve of degree $k$ to the points of $x[i]$ where $q[i] = 1$. The resulting curve can be specified using just $k$ elements. It perfectly reproduces $x[i]$ where $q[i] = 1$ and interpolates the remaining points.

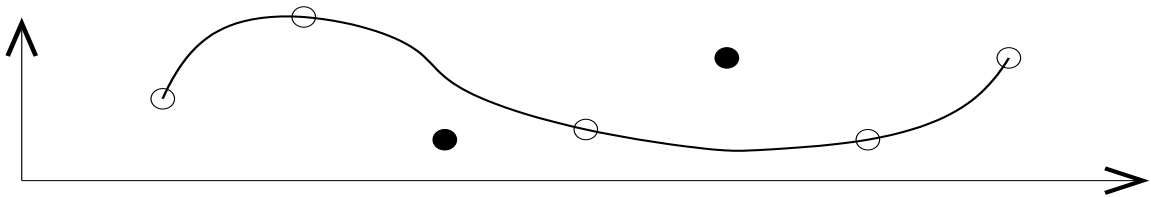

Fig. 2. Losslessly encoding a source with $n = 7$ points where only $k = 5$ points are relevant (*i.e.*, the unshaded ones), can be done by fitting a fourth degree curve to the relevant points. The resulting curve will require $k$ elements (yielding a compression ratio of $k/n$) and will exactly reproduce the desired points.

### B. Gaussian Source

A similar approach can be used to quantize a zero mean, unit variance, complex Gaussian source relative to quadratic distortion using the Discrete Fourier Transform (DFT). Specifically, to encode the source samples $x[0]$, $x[1]$, ..., $x[n-1]$, we view the $k$ relevant samples as elements of a complex, periodic, Gaussian, sequence with period $n$, which is band-limited in the sense that only its first $k$ DFT coefficients are non-zero. Using periodic, band-limited, interpolation we can use only the $k$ samples where $q[i] = 1$ to find the corresponding $k$ DFT coefficients $X[0]$, $X[1]$, ..., $X[k-1]$.

---

[2]The desired MDS code always exists since we assumed $|\mathcal{X}| \geq n$. For $|\mathcal{X}| < n$, near MDS codes exist, which give asymptotically similar performance with an overhead that goes to zero as $n \to \infty$.

The relationship between the $k$ relevant source samples and the $k$ interpolated DFT coefficients has a number of special properties. In particular this $k \times k$ transformation is unitary and furthermore each DFT basis vector has an equal amount of energy in each component of the original basis.[3] Hence, the DFT coefficients are Gaussian with unit variance and zero mean. Thus, the $k$ DFT coefficients can be quantized with average distortion $D$ per coefficient and $k \cdot R(D)$ bits where $R(D)$ represents the rate-distortion trade-off for the quantizer. To reconstruct the signal, the decoder simply transforms the quantized DFT coefficients back to the time domain. Since the DFT coefficients and the relevant source samples are related by a unitary transformation, the average error per coefficient for these source samples remains unchanged, *i.e.*, the error is exactly $D$.

Note if the side information were unavailable or ignored, then at least $n \cdot R(D)$ bits would be required. If the side information were losslessly sent to the decoder, then $n \cdot H_b(k/n) + k \cdot R(D)$ would be required. Finally, even if the decoder had knowledge of the side information, at least $k \cdot R(D)$ bits would be needed. Hence, the DFT scheme achieves the same performance as when the side information is available at both the encoder and decoder, and is strictly better than ignoring the side information or losslessly communicating it.

## III. NOTATION, PROBLEM MODEL, AND RATE-DISTORTION FUNCTIONS

Vectors and sequences are denoted in bold (*e.g.*, $\mathbf{x}$) with the $i$th element denoted as $x[i]$. Random variables are denoted using the sans serif font (*e.g.*, $\mathsf{x}$) while random vectors and sequences are denoted with bold sans serif (*e.g.*, $\mathbf{\mathsf{x}}$).

We are primarily interested in two kinds of side information, which we call "signal side information" and "distortion side information". The former (denoted $\mathbf{w}$) corresponds to information that is statistically related to the source but does not directly affect the distortion measure and the latter (denoted $\mathbf{q}$) corresponds to information that is not directly related to the source but does directly affect the distortion measure. Formally, we capture this decomposition with the following definition:

**Definition 1** *We define a source* $\mathbf{x}$*, distortion side information* $\mathbf{q}$*, signal side information* $\mathbf{w}$*, and distortion*

---

[3]The Hadamard transform as well as any permuted version of the DFT basis will also have these properties so the choice of transform is not unique. Simply choosing any unitary transform is not sufficient, however.

*measure*[4] $d(\mathbf{x}, \hat{\mathbf{x}}; \mathbf{q})$ *as an admissible side information decomposition if the following Markov chains are satisfied:*

$$\mathbf{q} \leftrightarrow \mathbf{w} \leftrightarrow \mathbf{x} \tag{2a}$$

$$d(\mathbf{x}, \hat{\mathbf{x}}; \mathbf{q}) \leftrightarrow \mathbf{x}, \hat{\mathbf{x}}, \mathbf{q} \leftrightarrow \mathbf{w}. \tag{2b}$$

The simplest case where these Markov chains are satisfied is when $\mathbf{q}$ and $\mathbf{w}$ are independent and therefore $\mathbf{q}$ and $\mathbf{x}$ are statistically independent (without conditioning on $\mathbf{w}$). While many of our results apply to the general setting, we often specialize our results to this case.

Decomposing side information into distortion related and signal related components proves useful since it allows us to isolate two important insights in source coding with side information. First, as Wyner and Ziv discovered [4], knowing $\mathbf{w}$ only at the decoder is often sufficient. Second, as our examples in Section II illustrate, knowing $\mathbf{q}$ only at the encoder is often sufficient. Furthermore, the relationship between the side information and the distortion measure and the relationship between the side information and the source often arise from physically different effects and so such a decomposition is warranted from a practical standpoint. Of course, such a decomposition is not always possible and we explore some issues for general side information $\mathbf{z}$, which affects both the source and distortion measure, in Sections V-C and VII-C.

In any case, we define the source coding with side information problem as the tuple

$$(\mathcal{X}, \hat{\mathcal{X}}, \mathcal{Q}, \mathcal{W}, p_\mathsf{x}(x), p_{\mathsf{w}|\mathsf{x}}(w|x), p_{\mathsf{q}|\mathsf{w}}(q|w), d(x, \hat{x}; q)). \tag{3}$$

Specifically, a source sequence $\mathbf{x}$ consists of the $n$ samples $\mathbf{x}[1], \mathbf{x}[2], \ldots, \mathbf{x}[n]$ drawn from the alphabet $\mathcal{X}$. The signal side information $\mathbf{w}$ and the distortion side information $\mathbf{q}$ likewise consist of $n$ samples drawn from the alphabets $\mathcal{W}$ and $\mathcal{Q}$ respectively. These random variables are generated according to the distribution

$$p_{\mathbf{x},\mathbf{q},\mathbf{w}}(\mathbf{x}, \mathbf{q}, \mathbf{w}) = \prod_{i=1}^{n} p_\mathsf{x}(x[i]) \cdot p_{\mathsf{w}|\mathsf{x}}(w[i]\,|\,x[i]) \cdot p_{\mathsf{q}|\mathsf{w}}(q[i]\,|\,w[i]). \tag{4}$$

Note that the distortion measure and joint distribution in (4) satisfy the admissibility condition of Definition 1 by construction.

---

[4]In (2b) we find it notationally convenient to consider the distortion measure as a random variable. This allows us to state the desired conditional independence relationship as a Markov chain, but in the paper we only consider standard distortion measures that are deterministic functions. This incurs no loss of generality since randomized distortion measures can always be replaced with expected values without changing our performance measures.

October 20, 2018 DRAFT

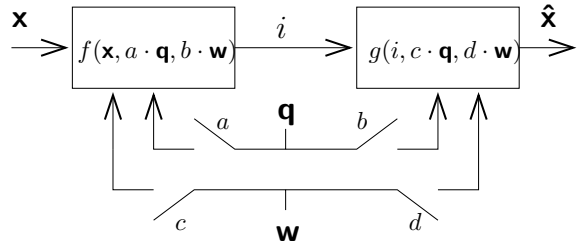

Fig. 3. Possible scenarios for source coding with distortion dependent and signal dependent side information **q** and **w**. The labels $a$, $b$, $c$, and $d$ are 0/1 if the corresponding switch is open/closed and the side information is unavailable/available to the encoder $f(\cdot)$ or decoder $g(\cdot)$.

Fig. 3 illustrates the sixteen possible scenarios where **q** and **w** may each be available at the encoder, decoder, both, or neither depending on whether the four switches are open or closed. A rate $R$ encoder $f(\cdot)$ maps a source as well as possible side information to an index $i \in \{1, 2, \ldots, 2^{nR}\}$. The corresponding decoder $g(\cdot)$ maps the resulting index as well as possible decoder side information to a reconstruction of the source. Distortion for a source **x**, which is quantized and reconstructed to the sequence $\hat{\mathbf{x}}$ taking values in the alphabet $\hat{\mathcal{X}}$, is measured via

$$d(\mathbf{x}, \hat{\mathbf{x}}; \mathbf{q}) = \frac{1}{n} \sum_{i=1}^{n} d(x[i], \hat{x}[i]; q[i]). \tag{5}$$

As usual, the rate-distortion function is the minimum rate such that there exists a system where the distortion is less than $D$ with probability approaching 1 as $n \to \infty$. We denote the sixteen possible rate-distortion functions by describing where the side-information is available. For example, $R[\text{Q-NONE-W-NONE}](D)$ denotes the rate-distortion function without side information and $R[\text{Q-NONE-W-DEC}](D)$ denotes the Wyner-Ziv rate-distortion function where **w** is available at the decoder [4]. Similarly, when all information is available at both encoder and decoder, $R[\text{Q-BOTH-W-BOTH}](D)$ describes Csiszár and Körner's [7] generalization of Gray's conditional rate-distortion function $R[\text{Q-NONE-W-BOTH}](D)$ [12] to the case where the side information can affect the distortion measure.

As pointed out by Berger [13], all the rate-distortion functions may be derived by considering **q** as part of **x** or **w** (*i.e.*, by considering the "super-source" $\mathbf{x}' = (\mathbf{x}, \mathbf{q})$ or the "super-side-information" $\mathbf{w}' = (\mathbf{w}, \mathbf{q})$) and applying well-known results for source coding, source coding with side information, the conditional rate-distortion theorem, *etc*. For example, if we set the signal side information **w** to null to simplify notation, the relevant rate-distortion functions are easy to obtain.

**Proposition 1** *The rate-distortion functions when* **w** *is null (and hence (3) implies* **x** *and* **q** *are independent) are*

$$R[\text{Q-NONE}](D) = \inf_{p_{\hat{x}|x}(\hat{x}|x) : E[d(x,\hat{x};q)] \leq D} I(x;\hat{x}) \tag{6a}$$

$$R[\text{Q-DEC}](D) = \inf_{p_{u|x}(u|x), v(\cdot,\cdot) : E[d(x,v(u,q);q)] \leq D} I(x;u) - I(u;q) \tag{6b}$$

$$R[\text{Q-ENC}](D) = \inf_{p_{\hat{x},q}(\hat{x}|x,q) : E[d(x,\hat{x};q)] \leq D} I(x,q;\hat{x}) = \inf_{p_{\hat{x},q}(\hat{x}|x,q) : E[d(x,\hat{x};q)] \leq D} I(x;\hat{x}|q) + I(\hat{x};q) \tag{6c}$$

$$R[\text{Q-BOTH}](D) = \inf_{p_{\hat{x},q}(\hat{x}|x,q) : E[d(x,\hat{x};q)] \leq D} I(x;\hat{x}|q). \tag{6d}$$

The rate-distortion functions in (6a), (6b), and (6d) follow from standard results (*e.g.*, [11], [6], [7], [12], [4]). To obtain (6c) we can apply the classical rate-distortion theorem to the "super source" $\mathbf{x}' = (\mathbf{x}, \mathbf{q})$.

## IV. WHEN ENCODER-ONLY KNOWLEDGE IS OPTIMAL

In this section, we consider the simplified case when the signal side information **w** is null (and hence Definition 1 implies **x** and **q** are independent) and derive conditions for when encoder-only knowledge of the distortion side information **q** is optimal. Comparing the rate-distortion functions in Proposition 1 immediately yields the following rate-loss result.

**Proposition 2** *Knowing* **q** *only at the encoder is as good as knowing it at both encoder and decoder if and only if there exists an $\hat{x}$ that optimizes (6d) with the property that $I(\hat{x};q) = 0$. In such a case, the resulting $\hat{x}$ is also optimal for (6c) and therefore (6c) and (6d) are equal.*

The intuition for the "only if" part of Proposition 2 is illustrated in Fig. 4. Specifically, $p_{\hat{x}|q}(\hat{x}|q)$ represents the distribution of the codebook. Thus if a different codebook is required for different values of $q$, then the penalty for knowing $q$ only at the encoder is exactly the information that the encoder must send to the decoder to allow the decoder to determine the proper codebook, *i.e.*, $I(\hat{x};q)$. The only way that knowing $q$ at the encoder can be just as good as knowing it at both is if there exists a fixed codebook that is universally optimal regardless of $q$.

One of the main insights of this paper is the intuition for the "if" part of Proposition 2: if a variable partition is allowed, universally good fixed codebooks exist as illustrated in Fig. 5. Specifically $p_{\hat{x}|q}(\hat{x}|q)$ represents the distribution of the codebook while $p_{\hat{x}|x,q}(\hat{x}|x,q)$ represents the quantizer partition mapping source and side information to a codeword. Thus even if the side information affects the distortion and $p_{\hat{x}|x,q}(\hat{x}|x,q)$ depends on $q$, it may be that $p_{\hat{x}|q}(\hat{x}|q)$ is independent of $q$. In such cases (characterized by the condition $I(\hat{x};q) = 0$), there exists a *fixed* codebook with a *variable* partition, which is simultaneously optimal for all values of the distortion side information **q**. Specifically, in such a system the reconstruction $\hat{\mathbf{x}}(i)$ corresponding to a particular index $i$ is fixed regardless of **q**, but the partition mapping **x** to the codebook index $i$ depends on **q**.

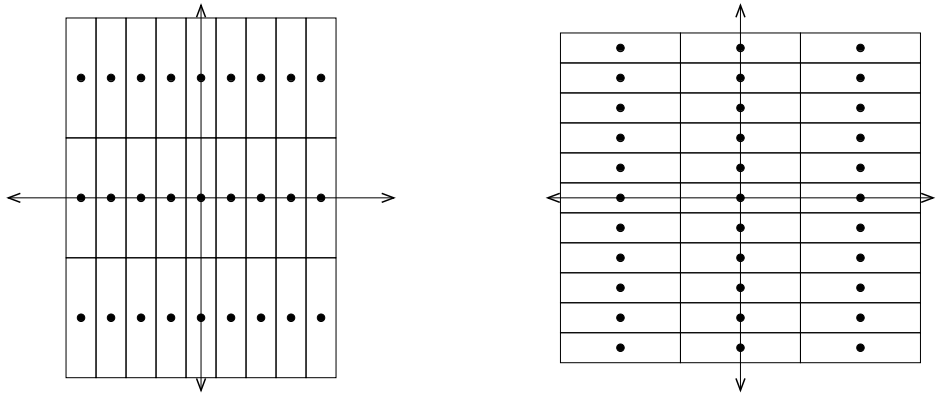

Fig. 4. An example of two different quantizer codebooks for different values of the distortion side information *q*. If *q* indicates the horizontal error (respectively, vertical error) is more important, then the encoder can use the codebook on the left (resp., right) to increase horizontal accuracy (resp., vertical accuracy). The penalty for knowing *q* only at the encoder is the amount of bits required to communicate which codebook was used.

In various scenarios, this type of fixed codebook variable partition approach can be implemented via a lattice [14] as illustrated in Fig. 6. As discussed in Section II, fixed codebook variable partition systems can also be constructed from transforms. Specifically, in Section II-B the fixed codebook is generated by quantizing bandlimited signals. The resulting variable partition has cells similar to Fig. 5 but with the width of the cells in the unimportant coordinates being infinite.

The quantization error will depend on the source distribution and size of the quantization cells. Thus if the quantization cells of a fixed codebook variable partition system like Fig. 5 are the same as the corresponding variable codebook system in Fig. 4, the performance will be the same. Intuitively, the main difference between these two figures (as well as general fixed codebook/variable partition schemes versus variable codebook schemes) is in the nature of the tiling of Fig. 5 versus Fig. 4. In Sections IV-A and Section IV-C, we consider various scenarios where the uniformity in the source or distortion measure makes these two tilings equivalent and thus the condition $I(\hat{x}; q) = 0$ is satisfied for all distortions. Similarly, in Section V we show that if we ignore "edge effects" and focus on the high-resolution regime, then the difference in these tilings becomes negligible. Thus in the high-resolution regime we show that a wide range of source and distortion models admit variable partition, fixed codebook quantizers as in

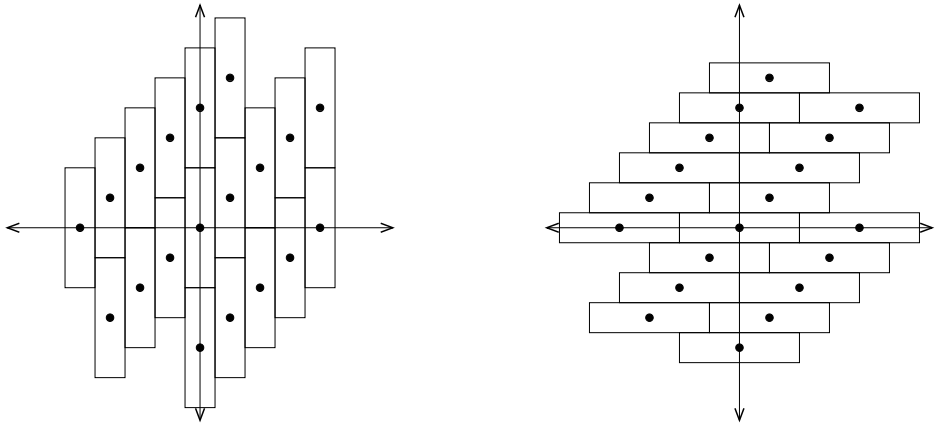

Fig. 5. An example of a quantizer with a variable partition and fixed codebook. If the encoder knows the horizontal error (respectively, vertical error) is more important, it can use the partition on the left (resp., right) to increase horizontal accuracy (resp., vertical accuracy). The decoder only needs to know the quantization point not the partition.

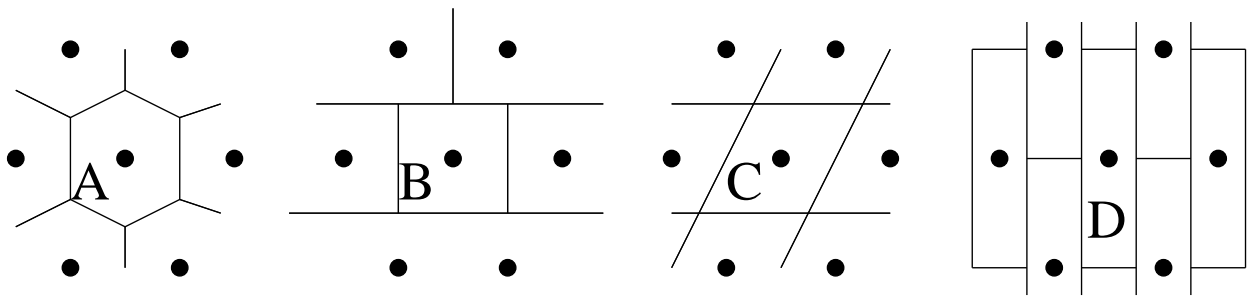

Fig. 6. A fixed, hexagonal, lattice codebook with four different partitions.

Fig. 5 and achieve $I(\hat{x}; q) = 0$.

Finally, we note that Proposition 2 suggests that distortion side information **q** complements signal side information **w** in the sense that **q** is useful at the encoder while **w** is useful at the decoder. In fact in Section V-B we strengthen this complementary relationship by showing that **w** is often useless at the encoder while **q** is often useless at the decoder.

### A. Uniform Sources with Group Difference Distortions

Let the source $x$ be uniformly distributed over a group $\mathcal{X}$ with the binary relation $\oplus$. For convenience, we use the symbol $a \ominus b$ to denote $a \oplus b^{-1}$ (where $b^{-1}$ denotes the additive inverse of $b$ in the group).

We define a group difference distortion measure as any distortion measure where

$$d(x, \hat{x}; q) = \rho(\hat{x} \ominus x; q) \tag{7}$$

for some function $\rho(\cdot; \cdot)$. As we will show, the symmetry in this scenario insures that the optimal codebook distribution is uniform. This allows an encoder to design a fixed codebook and vary the quantization partition based on **q** to achieve the same performance as a system where both encoder and decoder know **q**. This uniformity of the codebook, made precise in the following theorem, provides a general information theory explanation for the Reed-Solomon example in II-A.

**Theorem 1** *Let the source x be uniformly distributed over a group with a difference distortion measure and let the distortion side information q be independent of the source. Then, to optimize (6d), it suffices to choose the test-channel distribution $p_{\hat{x}|x,q}(\hat{x}|x,q)$ such that $p_{\hat{x}|q}(\hat{x}|q)$ is uniform for each q (and hence independent of q) implying that there is no rate-loss as stated in Proposition 2,* i.e.

$$R[\text{Q-ENC}](D) = R[\text{Q-BOTH}]. \tag{8}$$

For either finite or continuous groups this theorem can be proved by deriving the conditional Shannon Lower Bound (which holds for any source) and showing that this bound is tight for uniform sources. We use this approach below to give some intuition. For more general "mixed groups" with both discrete and continuous components, entropy is not well defined and a more complicated argument based on symmetry and convexity is provided in Appendix A.

**Lemma 1 (Conditional Shannon Lower Bound)** *Let the source x be distributed over a discrete group, $\mathcal{X}$, with a difference distortion measure, $\rho(x \ominus \hat{x}; q)$. Then if we define $z^*$ as the random variable that maximizes $H(z|q)$ subject to the constraint $E[\rho(z; q)] \leq D$, then*

$$R[\text{Q-ENC}](D) \geq \log |\mathcal{X}| - H(z^*|q). \tag{9}$$

*For continuous groups, $|\mathcal{X}|$ and $H(z^*|q)$ can be replaced by the Lebesgue measure of the group and $h(z^*|q)$ in (9) (as well as the following proof).*

*Proof:*

$$I(\hat{x}; x, q) \geq \log |\mathcal{X}| + H(q) - H(x, q|\hat{x}) \tag{10}$$

$$= \log |\mathcal{X}| + H(q) - H(q|\hat{x}) - H(\hat{x} \ominus x|x, q) \tag{11}$$

$$\geq \log |\mathcal{X}| - H(\hat{x} \ominus x|q) \tag{12}$$

$$\geq \log |\mathcal{X}| - H(z^*|q) \tag{13}$$

where (10) follows since a uniform source independent of $q$ maximizes entropy, (12) follows since conditioning reduces entropy, and (13) follows from the definition of $z^*$ since $E[\rho(\hat{x} \ominus x; q)] \leq D$. ∎

*Proof of Theorem 1:* Choosing the test-channel distribution $\hat{x} = z^* + x$ with the pair $(z^*, q)$ independent of $x$ achieves the bound in (9) with equality and must therefore be optimal. Furthermore, since $x$ is uniform, so is $\hat{x}$ and therefore $\hat{x}$ and $q$ are statistically independent. Therefore $I(\hat{x}; q) = 0$ and thus comparing (6c) to (6d) shows $R[\text{Q-ENC}](D) = R[\text{Q-BOTH}](D)$ for finite groups. The same argument holds for continuous groups with entropy replaced by differential entropy and $|\mathcal{X}|$ replaced by Lebesgue measure. ∎

## B. Examples

Uniform source and group difference distortion measures arise naturally in a variety of applications. One example is phase quantization where applications such as Magnetic Resonance Imaging, Synthetic Aperture Radar, and Ultrasonic Microscopy infer physical phenomena from the phase shifts induced in a probe signal [15], [16], [17]. Alternatively, when magnitude and phase information must both be recorded, there are sometimes advantages to treating these separately, [18], [19], [20], [21]. The key special case when only two phases are recorded corresponds to Hamming distortion and we use this scenario to illustrate how distortion side information affects quantization.

For a symmetric binary source, we first derive the various rate-distortion trade-offs for a general Hamming distortion measure depending on $q$. Next we adapt this general result to the special cases of quantizing noisy observations and quantizing with a weighted distortion measure. The former demonstrates that the naive encoding method where the encoder losslessly communicates the side information to the decoder and then uses optimal encoding, can require arbitrarily higher *rate* than the optimal rate-distortion trade-off. The second example demonstrates that ignoring the side information can result in arbitrarily higher *distortion* than the minimum required by optimal schemes.

*1) General Formula For Hamming Distortion Depending on* **q**: Consider a symmetric binary source *x* with side information $q$ taking values in $\{1, 2, \ldots, N\}$ with distribution $p_q(q)$. Let distortion be measured according to

$$d(x, \hat{x}; q) = \alpha_q + \beta_q \cdot d_H(x, \hat{x}) \tag{14}$$

where $\{\alpha_1, \alpha_2, \ldots, \alpha_N\}$ and $\{\beta_1, \beta_2, \ldots, \beta_N\}$ are sets of non-negative weights. In Appendix B we derive the various rate-distortion functions for $D \geq E[\alpha_q]$ to obtain

$$R[\text{Q-NONE}](D) = R[\text{Q-DEC}](D) = 1 - H_b\left(\frac{D - E[\alpha_q]}{E[\beta_q]}\right) \tag{15a}$$

$$R[\text{Q-ENC}](D) = R[\text{Q-BOTH}](D) = 1 - \sum_{i=1}^{N} p_q(i) \cdot H_b\left(\frac{2^{-\lambda \beta_i}}{1 + 2^{-\lambda \beta_i}}\right) \tag{15b}$$

where $\lambda$ is chosen to satisfy the distortion constraint

$$\sum_{i=1}^{N} p_q(i) \left[\alpha_i + \beta_i \cdot \frac{2^{-\lambda \beta_i}}{1 + 2^{-\lambda \beta_i}}\right] = D. \tag{16}$$

*2) Noisy Observations:* To provide a more concrete illustration of the effect of side information at the encoder, consider the special case where *x* is a noisy observation of an underlying source received through a binary symmetric channel (BSC) with cross over probability specified by the side information. Specifically, let the cross over probability of the BSC be

$$\epsilon_q = \frac{q-1}{2(N-1)} ,$$

which is always at most $1/2$.

A distortion of 1 is incurred if a cross over occurs due to either the noise in the observation or the noise in the quantization (but not both):

$$\begin{aligned} d(x, \hat{x}; q) &= \epsilon_q \cdot [1 - d_H(x, \hat{x})] + (1 - \epsilon_q) \cdot d_H(x, \hat{x}) \\ &= \epsilon_q + (1 - 2\epsilon_q) \cdot d_H(x, \hat{x}) \\ &= \frac{q-1}{2(N-1)} + \left(1 - \frac{q-1}{N-1}\right) \cdot d_H(x, \hat{x}). \end{aligned} \tag{17}$$

This corresponds to a distortion measure in the form of (14) with $\alpha_q = (q-1)/[2(N-1)]$ and $\beta_q = 1 - (q-1)/(N-1)$. Hence, the rate-distortion formulas from (15) apply. The optimal encoding strategy is to encode the noisy observation directly as discussed in [22] although with different amounts of quantization depending on the side information.

In the left plot of Fig. 7, we illustrate the rate-distortion function for this problem with and without side information at the encoder in the case where $q \in \{1, 2\}$ (*i.e.*, $N = 2$) and the observation is either noise-less or completely noisy. In the right plot of Fig. 7, we illustrate the rate-distortion functions in the limit where $N \to \infty$ and the noise is uniformly distributed between 0 and 1/2. Note that in the latter case, if the side information were encoded losslessly then $\log N$ extra bits of overhead would be required beyond the amount when optimal encoding is used. Hence, communicating the side information losslessly can require an arbitrarily large rate even though optimal use of the encoder side information reduces the quantization rate.

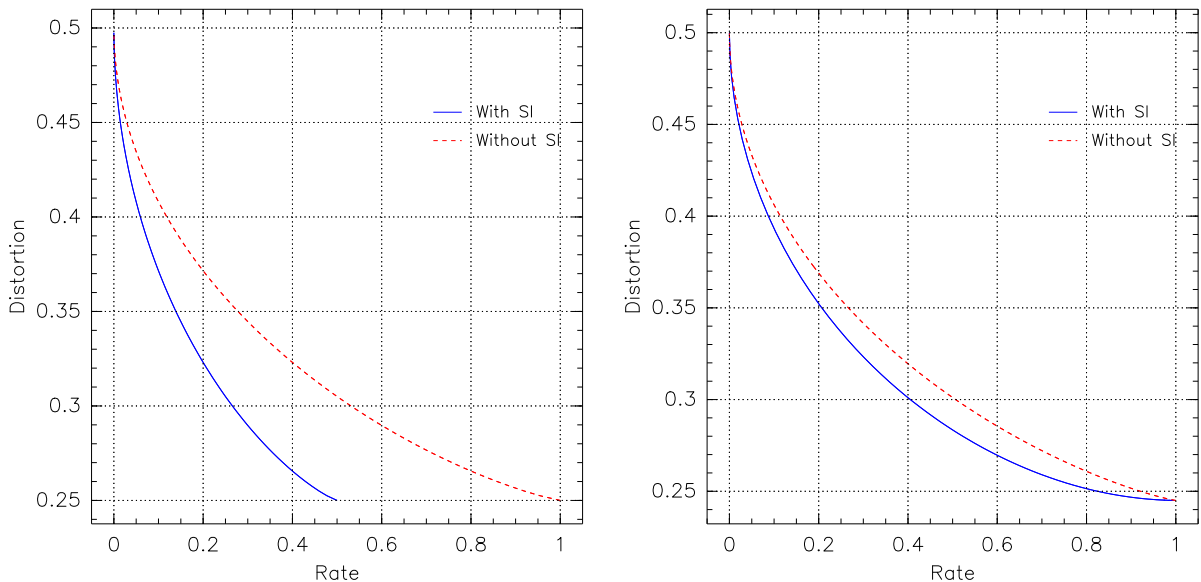

Fig. 7. Rate-Distortion curves for noisy observations of a binary source. The solid curve represents the minimum possible Hamming distortion when side information specifying the cross-over probability of the observation noise is available at the encoder (or both at the encoder and decoder). The dashed curve represents the minimum distortion when side information is not available (or ignored) at the encoder. For the plot on the left the cross over probability for the observation noise is equally likely to be 0 or 1/2, while for the plot on the right it is uniformly distributed over the interval $[0, 1/2]$.

*3) Weighted Distortion:* In the previous example, certain source samples were more important because they were observed with less noise. In this section we consider a model where certain samples of a source are inherently more important than others (*e.g.*, edges in a binary image, other perceptually important features, or sensor readings in high activity areas). Specifically, we consider a distortion measure of the form (14) where $\beta_k = \exp(\gamma k/N)$, $\alpha_k = 0$, and the side information is uniformly distributed over $\{0, 1, \ldots, N-1\}$.

The left plot in Fig. 8 illustrates the rate-distortion curves for the case when $N = 2$, while the right plot corresponds to the case where $N \to \infty$. The former model corresponds to two types of samples:

important samples where a distortion of $\exp(\gamma/2)$ is incurred if quantized incorrectly and normal samples where a distortion of 1 is incurred if quantized incorrectly. If no side information is available (or if side information is ignored), the encoder must treat these samples equally. If side information at the encoder is used optimally, then more bits are spent quantizing the important samples. The latter model illustrates the case when there is a continuum of importance for the samples.

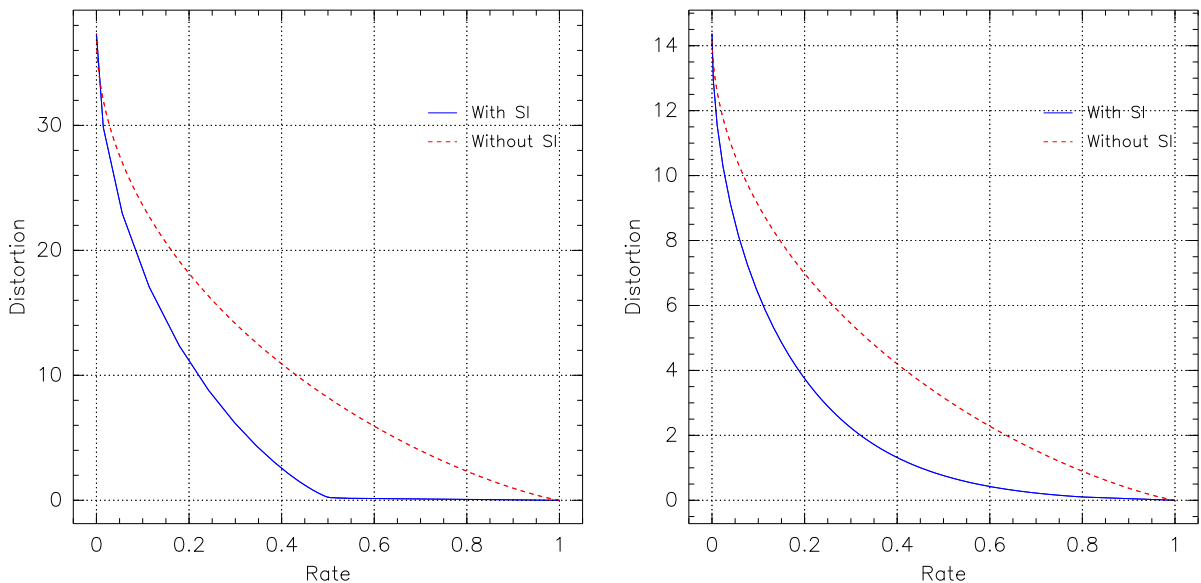

Fig. 8. Rate-Distortion curves for a binary source with weighted Hamming distortion. The distortion for quantizing each source sample is measured via Hamming distortion times the weight $\exp(5 \cdot q)$. The solid curve represents the minimum possible Hamming distortion when side information specifying the weight is available at the encoder (or both at the encoder and decoder). The dashed curve represents the minimum distortion when no side information is available at the encoder. In the left plot, $q$ is uniformly distributed over the pair $\{0, 1\}$ while in the right plot $q$ is uniformly distributed over the interval $[0, 1]$.

In the limit as $\gamma \to \infty$ and $N = 2$, the system not using side information, suffers increasingly more distortion. This is most evident for rates greater than 1/2. In this rate region, the system with side information losslessly encodes the important samples and the distortion is bounded by 0.5 while the system without side information has a distortion that scales with $\exp(\gamma/2)$. Thus the extra distortion incurred when $q$ is not available to the encoder can be arbitrarily large.

## C. Erasure Distortions

Another scenario where the condition $I(\hat{x}; q) = 0$ is satisfied is for "erasure distortions" where $q$ is binary and the distortion measure is of the form

$$d(x, \hat{x}; q) = \rho(x, \hat{x}) \cdot q \tag{18}$$

for some function $\rho(\cdot, \cdot)$.

**Theorem 2** *For any source distribution, if the distortion measure is of the form in* (18) *with binary* **q***, then the rate-distortion function when* **q** *is known at only the encoder is the same as when* **q** *is known at both encoder and decoder,* i.e.,

$$R[\text{Q-ENC}](D) = R[\text{Q-BOTH}](D). \tag{19}$$

*Proof:* Let $\hat{x}^*$ be a distribution that optimizes (6d). Choose the new random variable $\hat{x}^{**}$ to be the same as $\hat{x}^*$ when $q = 0$ and when $q = 1$, let $\hat{x}^{**}$ be independent of $x$ with the same marginal distribution as when $q = 0$:

$$p_{\hat{x}^{**}|x,q}(\hat{x}|x, q) = \begin{cases} p_{\hat{x}^*|x,q}(\hat{x}|x, q), & q = 0 \\ p_{\hat{x}^*|q}(\hat{x}|q = 0), & q = 1. \end{cases} \tag{20}$$

Both $\hat{x}^*$ and $\hat{x}^{**}$ have the same expected distortion since they only differ when $q = 0$. Furthermore, by the data processing inequality

$$I(\hat{x}^{**}; x|q) \leq I(\hat{x}^*; x|q) \tag{21}$$

so $\hat{x}^{**}$ also optimizes (6d). Finally, since $I(\hat{x}^{**}; q) = 0$, Proposition 2 is satisfied and we obtain the desired result. ∎

As shown in the examples of Section II, erasure distortions may admit particularly simple quantizers to optimally use encoder side information.

## V. Asymptotically Zero Loss for High-Resolution Quantization

In Proposition 2 of Section IV we saw that knowing **q** only at the encoder is as good as full knowledge of **q** when the codebook is independent of the distortion side information (*i.e.*, when $I(\hat{x}; q) = 0$). In the high-resolution regime,[5] $\hat{\mathbf{x}}$ begins to closely approximate **x**. Thus when **x** and **q** are independent we may intuitively expect that as $\hat{\mathbf{x}} \to \mathbf{x}$ we have $I(\hat{x}; q) \to I(x; q) = 0$. In this section, we rigorously justify this intuition and also consider some variations.

---

[5]Usually, the high-resolution limit is defined as when $D \to 0$, but it is sometimes useful to consider distortion measures with a constant penalty. Hence we assume it is possible to define a minimum distortion $D_{\min}$ which can be approached arbitrarily closely as the rate increases and we define the high-resolution limit as $D \to D_{\min}$.

For example, we consider the more general case where the signal side information **w**, which is statistically dependent on the source, is present. Our problem model (and specifically the Markov conditions in (2)) require **q** to be conditionally independent of **x** given **w** and require the distortion to be conditionally independent of **w** given **q**, **x**, and $\hat{\mathbf{x}}$. But since our model allows for **q** and **w** to be statistically dependent, **q** may now be indirectly correlated with **x** (through **w**) and **w** may indirectly affect the distortion (through **q**). Even in this more general scenario, the basic intuition that distortion side information is necessary only at the encoder and signal side information is necessary only at the decoder continues to hold in many cases of interest.

## A. Technical Conditions

In addition to the side information decomposition implied by Definition 1, our results require a "continuity of entropy" property that essentially states

$$z \to 0 \Rightarrow h(x + z | q, w) \to h(x | q, w). \tag{22}$$

The desired continuity follows from [23] provided the source, distortion measure, and side information satisfy some technical conditions related to smoothness. These conditions are not particularly hard to satisfy; for example, any vector source, side information, and distortion measure where

$$\exists \delta > 0, -\infty < E[||\mathbf{x}||^\delta | \mathbf{w} = \mathbf{w}] < \infty \quad \forall \mathbf{w} \tag{23a}$$

$$-\infty < h(\mathbf{x}|\mathbf{w} = \mathbf{w}) < \infty, \quad \forall \mathbf{w} \tag{23b}$$

$$d(\mathbf{x}, \hat{\mathbf{x}}; \mathbf{q}) = \alpha(\mathbf{q}) + \beta(\mathbf{q}) \cdot ||\mathbf{x} - \hat{\mathbf{x}}||^{\gamma(\mathbf{q})} \tag{23c}$$

will satisfy the desired technical conditions in [23] provided $\alpha(\cdot)$, $\beta(\cdot)$, and $\gamma(\cdot)$ are non-negative functions. For more general scenarios we introduce the following definition to summarize the requirements from [23].

**Definition 2** *We define a source x, side information pair (q, w), and difference distortion measure $d(x, \hat{x}; q) = \rho(x - \hat{x}; q)$ as admissible if the following conditions are satisfied:*

1) *the equations*

$$a(D, q) \int \exp[-s(D, q)\rho(x; q)] dx = 1 \tag{24a}$$

$$a(D, q) \int \rho(x; q) \exp[-s(D, q)\rho(x; q)] dx = D \tag{24b}$$

*have a unique pair of solutions $(a(D, q), s(D, q))$ for all $D > D_{\min}$ that are continuous functions of their arguments*

2) $-\infty < h(\mathsf{x}|\mathsf{w} = w) < \infty$, *for all $w$*

3) *For each value of $q$, there exists an auxiliary distortion measure $\delta(\cdot; q)$ where the equations*

$$a_\delta(D, q) \int \exp[-s_\delta(D, q)\delta(x; q)]dx = 1 \tag{25a}$$

$$a_\delta(D, q) \int \delta(x; q) \exp[-s_\delta(D, q)\delta(x; q)]dx = D \tag{25b}$$

*have a unique pair of solutions $(a_\delta(D, q), s_\delta(D, q))$ for all $D > D_{\min}$ that are continuous functions of their arguments*

4) *The distribution $\mathsf{z_q}$ that maximizes $h(\mathsf{z_q}|\mathsf{q})$ subject to the constraint $E[\rho(\mathsf{z_q}; \mathsf{q})] \leq D$ has the property that*

$$\lim_{D \to D_{\min}} \mathsf{z_q} \to 0 \text{ in distribution } \forall q \tag{26a}$$

$$\lim_{D \to D_{\min}} E[\delta(\mathsf{x} + \mathsf{z_q}, \mathsf{q})|\mathsf{q} = q] = E[\delta(\mathsf{x}, \mathsf{q})|\mathsf{q} = q] \quad \forall q. \tag{26b}$$

### B. Equivalence Theorems

Our main results for continuous sources consist of various theorems describing when different types of side information knowledge are equivalent. All our results require the basic side information decomposition and Markov chains in Definition 1. Some results require further conditions such as high-resolution, scaled difference distortion measures, or statistical independence between **q** and **w** (*e.g.*, when **w** is known only at the decoder, knowing **q** only at the encoder is asymptotically as good as knowing it at both when **q** and **w** are independent). The equivalence theorems stated below are proved in Appendix C and summarized in Fig. 9. Essentially, we show that the sixteen possible information configurations can be reduced to the four shown in Fig. 9. Specifically, to determine the performance of a given side information configuration one can essentially ask two questions: "Does the encoder have *q*?" and "Does the decoder have *w*?". A negative answer to either question incurs some penalty relative to the case with all side information known everywhere. In contrast, a positive answer to both questions incurs asymptotically no rate-loss relative to complete side information.

We begin by generalizing previous results about no rate-loss for the Wyner-Ziv problem in the high-resolution limit [8] [10] to show there is no rate-loss when **q** is known at the encoder and **w** is known at the decoder. This results suggests that there is a natural division of side information between the

October 20, 2018　　　　　　　　　　　　　　　　　　　　　　　　　　　　　　　　　　　　　　　　　　　　　　　　　　　　　　　　　　　DRAFT

|  | Decoder missing **w** | | Decoder has **w** | |
|---|---|---|---|---|
| Encoder Missing **q** | $R[\text{Q-DEC-W-ENC}]$ $\overset{\text{Th. 4 (I)}}{\Longleftrightarrow}$ $R[\text{Q-DEC-W-NONE}]$ | | $R[\text{Q-DEC-W-BOTH}]$ $\overset{\text{Th. 5 (S)}}{\Longleftrightarrow}$ $R[\text{Q-NONE-W-BOTH}]$ | |
|  | $\Updownarrow$ Th. 5 (S+I) | $\Updownarrow$ Th. 5 (S+I) | $\Updownarrow$ Th. 6 (H+S+I) | $\Updownarrow$ Th. 6 (H+S+I) |
|  | $R[\text{Q-NONE-W-ENC}]$ $\overset{\text{Th. 4 (I)}}{\Longleftrightarrow}$ $R[\text{Q-NONE-W-NONE}]$ | | $R[\text{Q-DEC-W-DEC}]$ $\overset{\text{Th. 5 (S)}}{\Longleftrightarrow}$ $R[\text{Q-NONE-W-DEC}]$ | |
| Encoder has **q** | $R[\text{Q-ENC-W-ENC}]$ $\overset{\text{Th. 4}}{\Longleftrightarrow}$ $R[\text{Q-ENC-W-NONE}]$ | | $R[\text{Q-ENC-W-DEC}]$ $\overset{\text{Th. 6 (H)}}{\Longleftrightarrow}$ $R[\text{Q-ENC-W-BOTH}]$ | |
|  | $\Updownarrow$ Th. 6 (H+I) | $\Updownarrow$ Th. 6 (H+I) | $\Updownarrow$ Th. 6 (H) | $\Updownarrow$ Th. 6 (H) |
|  | $R[\text{Q-BOTH-W-ENC}]$ $\overset{\text{Th. 4}}{\Longleftrightarrow}$ $R[\text{Q-BOTH-W-NONE}]$ | | $R[\text{Q-BOTH-W-DEC}]$ $\overset{\text{Th. 6 (H)}}{\Longleftrightarrow}$ $R[\text{Q-BOTH-W-BOTH}]$ | |

Fig. 9. Summary of main results for continuous sources. Arrows indicate which theorems demonstrate equality between various rate-distortion functions and list the assumptions required (H = high-resolution, I = **q** and **w** independent, S = scaled difference distortion).

encoder and decoder (at least asymptotically). Specifically, distortion side information should be sent to the encoder and signal side information should be sent to the decoder.

**Theorem 3** *For any admissible source, side information, and difference distortion measure satisfying Definitions 1 and 2,* **q** *and* **w** *can be divided between the encoder and decoder with no asymptotic penalty,* i.e.,

$$\lim_{D \to D_{\min}} R[\text{Q-ENC-W-DEC}](D) - R[\text{Q-BOTH-W-BOTH}](D) = 0. \tag{27}$$

The next two theorems generalize Berger's result that signal side information is useless when known only at the encoder [11] to show when **w** and **q** are useless at the encoder and decoder respectively. These results suggest that deviating from the natural division of Theorem 3 and providing side information in the wrong place makes that side information useless (at least in terms of the rate-distortion function).

**Theorem 4** *Let* **q** *and* **w** *be independent[6] and consider a difference distortion measure of the form $d(x - \hat{x}; q)$. Then* **w** *provides no benefit when known only at the encoder,* i.e.,

$$R[\text{Q-*-W-ENC}](D) = R[\text{Q-*-W-NONE}](D) \tag{28}$$

*where the wildcard "*" may be replaced with an element from* {ENC, DEC, BOTH, NONE} *(both *'s must be replaced with the same element).*

---

[6]Independence is only required when $* \in \{\text{DEC}, \text{NONE}\}$; if $* \in \{\text{ENC}, \text{BOTH}\}$, the theorem holds without this condition provided the side information decomposition is admissible according to Definition 1.

**Theorem 5** *Let* **q** *and* **w** *be independent*[7] *and consider a scaled distortion measure of the form* $d(x, \hat{x}; q) = d_0(q) d_1(x, \hat{x})$. *Then* **q** *provides no benefit when known only at the decoder,* i.e.,

$$R[\text{Q-DEC-W-*}](D) = R[\text{Q-NONE-W-*}](D) \tag{29}$$

*where the wildcard "*" may be replaced with an element from* {ENC, DEC, BOTH, NONE} *(both *'s must be replaced with the same element).*

Using the previous results, we can generalize Theorem 3 to show that regardless of where **w** (respectively **q**) is known, knowing **q** (**w**) only at the encoder (decoder) is sufficient in the high-resolution limit. This result essentially suggests that even if the ideal of providing **q** to the encoder and **w** to the decoder suggested by Theorem 3 is impossible, it is still useful to follow this ideal as much as possible.

**Theorem 6** *Let* **q** *and* **w** *be independent.*[8] *For any source and scaled*[9] *difference distortion measure* $d(x, \hat{x}; q) = d_0(q) \cdot d_1(x - \hat{x})$ *satisfying the conditions in Definitions 1 and 2,* **q** *(respectively,* **w***) is asymptotically only required at the encoder (respectively, at the decoder),* i.e.,

$$\lim_{D \to D_{\min}} R[\text{Q-ENC-W-*}](D) - R[\text{Q-BOTH-W-*}](D) = 0 \tag{30a}$$

$$\lim_{D \to D_{\min}} R[\text{Q-*-W-DEC}](D) - R[\text{Q-*-W-BOTH}](D) = 0 \tag{30b}$$

*where the wildcard "*" may be replaced with an element from* {ENC, DEC, BOTH, NONE} *(both *'s must be replaced with the same element).*

## C. Penalty Theorems

We can compute the asymptotic penalty for not knowing the signal side information **w** at the decoder.

---

[7] Independence is only required when $* \in \{\text{ENC}, \text{NONE}\}$; if $* \in \{\text{DEC}, \text{BOTH}\}$, the theorem holds without this condition provided the side information decomposition is admissible according to Definition 1.

[8] Independence is only required when $* \in \{\text{ENC}, \text{NONE}\}$ in (30a) or when $* \in \{\text{DEC}, \text{NONE}\}$ in (30b). For $* \in \{\text{DEC}, \text{BOTH}\}$ in (30a) or $* \in \{\text{ENC}, \text{BOTH}\}$ in (30b) the theorem holds without this condition provided the side information decomposition is admissible according to Definition 1.

[9] The scaled form of the distortion measure is only required when $* \in \{\text{DEC}, \text{NONE}\}$ in (30b). When $* \in \{\text{ENC}, \text{BOTH}\}$, the theorem only requires a difference distortion measure of the form $d(x, \hat{x}; q) = d'(x - \hat{x}; q)$.

**Theorem 7** *Let **q** and **w** be independent.*[10] *Then for any source and scaled difference distortion measure $d(x, \hat{x}; q) = d_0(q) \cdot d_1(x - \hat{x})$ satisfying the conditions in Definition 2, the penalty for not knowing **w** at the decoder is*

$$\lim_{D \to D_{\min}} R[\text{Q-*-W-\{ENC-OR-NONE\}}](D) - R[\text{Q-*-W-\{DEC-OR-BOTH\}}](D) = I(x; w) \tag{31}$$

*where the wildcard "\*" may be replaced with an element from* {ENC, DEC, BOTH, NONE} *(all \*'s must be replaced with the same element).*

Theorem 7 also gives us insight about distortion side information that is not independent of the source. Specifically, imagine that the side information z affects the distortion via $d(x, \hat{x}; z) = d_0(z) \cdot d_1(x - \hat{x})$ and furthermore, z is correlated with the source. What is the penalty for knowing z only at the encoder versus at both encoder and decoder? To answer this question, we can decompose z into our framework by setting q = z and w = z and computing

$$\lim_{D \to D_{\min}} R[\text{Q-ENC-W-ENC}](D) - R[\text{Q-BOTH-W-BOTH}](D). \tag{32}$$

Since this pair of q and w are statistically dependent, we take into account the footnote in Theorem 7. Therefore we see that the asymptotic penalty for knowing general side information only at the encoder is exactly the degree to which the source and distortion side information are related as measured by mutual information:

**Corollary 1** *For any source and scaled difference distortion measure $d(x, \hat{x}; z) = d_0(z) \cdot d_1(x - \hat{x})$ satisfying the conditions in Definition 2, the penalty for knowing general side information z only at the encoder is*

$$\lim_{D \to D_{\min}} R[\text{Z-ENC}](D) - R[\text{Z-BOTH}](D) = I(x; z). \tag{33}$$

Finally, we can compute the asymptotic penalty for not knowing the distortion side information **q** at the encoder.

**Theorem 8** *Let **q** and **w** be independent.*[11] *For any source taking values in the $k$-dimensional real vector space and a scaled, norm-based distortion measure $d(\mathbf{x}, \hat{\mathbf{x}}; \mathbf{q}) = \mathbf{q} \cdot ||\mathbf{x} - \hat{\mathbf{x}}||^r$ satisfying the conditions*

---

[10]Independence is only required when $* \in \{\text{DEC}, \text{NONE}\}$; if $* \in \{\text{ENC}, \text{BOTH}\}$, the theorem holds without this condition provided the side information decomposition is admissible according to Definition 1.

[11]Independence is only required when $* \in \{\text{ENC}, \text{NONE}\}$; if $* \in \{\text{DEC}, \text{BOTH}\}$, the theorem holds without this condition provided the side information decomposition is admissible according to Definition 1.

*in Definition 2, the penalty (in nats/sample) for not knowing* **q** *at the encoder is*

$$\lim_{D \to D_{\min}} R[\text{Q-\{DEC-OR-NONE\}-W-*}](D) - R[\text{Q-\{ENC-OR-BOTH\}-W-*}](D) = \frac{k}{r} E\left[\ln \frac{E[q]}{q}\right] \quad (34)$$

*where the wildcard "*" may be replaced with an element from* {ENC, DEC, BOTH, NONE} *(both *'s must be replaced with the same element).*

A similar result, which essentially compares the asymptotic difference between

$$R[\text{Q-DEC-W-DEC}](D) \text{ and } R[\text{Q-BOTH-W-BOTH}](D)$$

for non-independent *q* and *w* with squared norm distortion, is derived in [10]. Thus as with Corollary 1, [10] and Theorem 8 can be interpreted as saying that the asymptotic penalty for knowing general side information *z* only at the decoder can be quantified by the degree to which the distortion and the side information are related as measured by (34). Specifically, setting *q = z* and *w = z* yields distortion side information and signal side information that are statistically dependent. Thus from [10] or by applying the footnote to Theorem 8 we obtain the following Corollary:

**Corollary 2** *For any source taking values in the $k$-dimensional real vector space and a scaled, norm-based distortion measure $d(\mathbf{x}, \hat{\mathbf{x}}; z) = z \cdot ||\mathbf{x} - \hat{\mathbf{x}}||^r$ satisfying the conditions in Definition 2, the penalty (in nats/sample) for not knowing z at the encoder is*

$$\lim_{D \to D_{\min}} R[\text{Z-DEC}](D) - R[\text{Z-BOTH}](D) = \frac{k}{r} E\left[\ln \frac{E[z]}{z}\right]. \quad (35)$$

According to Jensen's inequality, this rate gap is always greater than or equal to zero with equality if and only if the distortion side information is a constant with probability 1. Furthermore, the rate gap is scale invariant in the sense that it does not change when the distortion side information is multiplied by any positive constant.

In Table I, we evaluate the high-resolution rate penalty for a number of possible distributions for the side-information. Note that for all of these side information distributions (except the uniform and exponential distributions), the rate penalty can be made arbitrarily large by choosing the appropriate shape parameter to place more probability near $q = 0$ or $q = \infty$. In the former case (LogNormal, Gamma, or Pathological *q*), the large rate-loss occurs because when $q \approx 0$, the informed encoder can transmit almost zero rate while the uninformed encoder must transmit a large rate to achieve high resolution. In the latter case (Pareto or Cauchy *q*), the large rate-loss is caused by the heavy tails of the distribution for *q*. Specifically, even though *q* is big only very rarely, it is the rare samples of large *q* that dominate the moments. Thus an informed encoder can describe the source extremely accurately during the rare occasions when *q* is large, while an uninformed encoder must always spend a large rate to obtain a low average distortion.

TABLE I

ASYMPTOTIC RATE-PENALTY IN NATS. EULER'S CONSTANT IS DENOTED BY $\gamma$.

The rate penalties below are for not knowing distortion side information *q* at the encoder when distortion is measured via $d(x, \hat{x}; q) = q(x - \hat{x})^2$. (Multiply penalties in nats by $1/\ln 2 \approx 1.44$ to convert to bits).

| Distribution Name | Density for *q* | Rate Gap in nats |
|---|---|---|
| Exponential | $\tau \exp(-q\tau)$ | $-\frac{1}{2} \ln \gamma \approx 0.2748$ |
| Uniform | $1_{q \in [0,1]}$ | $\frac{1}{2}(1 - \ln 2) \approx 0.1534$ |
| Lognormal | $\frac{1}{q\sqrt{2\pi Q^2}} \exp\left[-\frac{(\ln q - M)^2}{2Q^2}\right]$ | $\frac{Q^2}{4}$ |
| Pareto | $\frac{a^b}{q^{a+1}}, q \geq b > 0, a > 1$ | $\frac{1}{2}\left[\ln \frac{a}{a-1} - 1/a\right]$ |
| Gamma | $\frac{b(bq)^{a-1} \exp(-bq)}{\Gamma(a)}$ | $\frac{1}{2}\left\{\ln a - \frac{d}{dx}[\ln \Gamma(x)]_{x=a}\right\} \approx \frac{1}{2a}$ |
| Pathological | $(1-\epsilon)\delta(q - \epsilon) + \epsilon\delta(q - 1/\epsilon)$ | $\frac{1}{2}\ln(1 + \epsilon - \epsilon^2) - \frac{1-2\epsilon}{2}\ln \epsilon \approx \frac{1}{2}\ln\frac{1}{\epsilon}$ |
| Positive Cauchy | $\frac{2/\pi}{1+q^2}, q \geq 0$ | $\infty$ |

Finally, note that all but one of these distributions would require infinite rate to losslessly communicate the side information. Thus the gains in using distortion side information *cannot* be obtained by exactly describing the side information to the receiver.

## VI. NON-ASYMPTOTIC BOUNDS FOR QUADRATIC DISTORTIONS

So far we have shown that in the high-resolution limit, knowing **q** at the encoder is sufficient. Our main analytical tool was the additive test-channel distribution $\hat{x} = x + z_q$ where the additive noise in the test channel depends on the distortion side information. Evidently, additive noise test-channels of this type are asymptotically optimal. To investigate the rate-loss at finite resolutions we develop two results for scaled quadratic distortion measures. We also briefly mention how these results can be generalized to other distortion measures.

## A. A Medium Resolution Bound Using Fisher Information

**Theorem 9** *Consider a scaled quadratic distortion measure of the form $d(x, \hat{x}; q) = q \cdot (x - \hat{x})^2$ with $q \geq q_{\min} > 0$. Then the maximum rate-gap at distortion $D$ is bounded by*

$$R[\text{Q-ENC-W-DEC}](D) - R[\text{Q-BOTH-W-BOTH}](D) \leq \frac{J(x|w)}{2} \cdot \min\left[1, \frac{D}{q_{\min}}\right] \quad (36)$$

*where $J(x|w)$ is the Fisher Information in estimating a non-random parameter $\tau$ from $\tau + x$ conditioned on knowing $w$. Specifically,*

$$J(x|w) \triangleq \int p_w(w) \left\{ \int p_{x|w}(x|w) \left[\frac{\partial}{\partial x} \log p_{x|w}(x|w)\right]^2 dx \right\} dw. \quad (37)$$

Similar bounds can be developed with other distortion measures provided that $D/q_{\min}$ is replaced with a quantity proportional to the variance of the quantization error. See the remark after the proof of Theorem 9 in the appendix for details. Also, Zamir and Feder discuss related bounds in [24, Appendix D].

One may wonder why Fisher Information appears in the above rate-loss bound. After all, Fisher Information is most commonly used to lower bound the error in estimating a parameter via its use in the Cramer-Rao bound. What does Fisher Information have to do with source coding?

To answer this question, recall that our bounds are all developed by using an additive test-channel distribution of the form $\hat{x} = x + z$. Thus, a clever source decoder could treat each source sample $x[i]$ as a parameter to be "estimated" from the quantized representation $\hat{x}[i]$. If an efficient estimator exists, this procedure could potentially reduce the distortion by the reciprocal of the Fisher Information. But if the distortion can be reduced in this manner without affecting the rate, then the additive test-channel distribution must be sub-optimal and a rate gap must exist.

So the bound in Theorem 9 essentially measures the rate gap by measuring how much our additive test-channel distribution could be improved if an efficient estimator existed for $x$ given $\hat{x}$. This bound will tend to be good when an efficient estimator does exist and poor otherwise. For example, if $x$ is Gaussian with unit-variance conditioned on $w$, then the Fisher Information term in (36) evaluates to one and the worst-case rate-loss is at most half a bit at maximum distortion. This corresponds to the half-bit bound on the rate-loss for the pure Wyner-Ziv problem derived in [8]. But if $x$ is discontinuous (*e.g.*, if $x$ is uniform), then no efficient estimator exists and the bound in (36) is poor.

As an aside, we note that the proof of Theorem 9 does not require any extra regularity conditions. Hence, if the Fisher Information of the source is finite, it can be immediately applied without the need to check whether the source is admissible according to Definition 2.

## B. A Low Resolution Bound

While the Fisher Information bound from (36) can be used at low resolutions, it can be quite poor if the source is not smooth. Therefore, we derive the following bound on the rate-loss, which is independent of the distortion level and hence most useful at low resolution.

**Theorem 10** *Consider a scaled quadratic distortion measure of the form $d(x, \hat{x}; q) = q \cdot (x - \hat{x})^2$ and denote the minimum/maximum conditional variance of the source by*

$$\sigma^2_{\min} = \min_w \text{Var}\,[x|w = w] \tag{38a}$$

$$\sigma^2_{\max} = \max_w \text{Var}\,[x|w = w]. \tag{38b}$$

*Then the gap in Theorem 3 is at most the conditional relative entropy of the source from a Gaussian distribution plus a term depending on the range of the conditional variance:*

$$R[\text{Q-ENC-W-DEC}](D) - R[\text{Q-BOTH-W-BOTH}](D) \leq D(p_{x|w}||\mathcal{N}(\text{Var}\,[x])) + \frac{1}{2}\log\left(1 + \frac{\sigma^2_{\max}}{\sigma^2_{\min}}\right) \tag{39}$$

*where $\mathcal{N}(t)$ represents a Gaussian random variable with mean zero and variance $t$.*

Similar bounds can be developed for other distortion measures as discussed after the proof of Theorem 10 in the appendix.

Consider the familiar Wyner-Ziv scenario where the signal side information is a noisy observation of the source. Specifically, let $w = x + v$ where $v$ is independent of $x$. In this case, the conditional variance is constant and (39) becomes

$$D(p_{x|w}||\mathcal{N}(\text{Var}\,[x])) + \frac{1}{2}\log 2 \tag{40}$$

and the rate-loss is at most half a bit plus the deviation from Gaussianity of the source.

If $x$ is Gaussian when conditioned on $w = w$, then the rate-loss is again seen to be at most half a bit as in [8]. In contrast to [8], which is independent of the source, however, both our bounds in (36) and (39) depend on the source distribution. Hence, we conjecture that our bounds are loose. In particular, for a discrete source, the worst case rate loss is at most $H(x|w)$, but this is not captured by our results since both bounds become infinity. Techniques from [25], [26], [8] may yield tighter bounds.

## C. A Finite Rate Gaussian Example

To gain some idea of when the asymptotic results take effect, we consider a finite rate Gaussian example. Specifically, let the source consist of a sequence of Gaussian random variables with mean zero and variance 1 and consider distortion side information with $\Pr[q = 1] = 0.6$, $\Pr[q = 10] = 0.4$, and distortion measure $d(x, \hat{x}; q) = q \cdot (x - \hat{x})^2$.

The case without side information is equivalent to quantizing a Gaussian random variable with distortion measure $4.6(x - \hat{x})^2$ and thus the rate-distortion function is

$$R[\text{Q-NONE-W-NONE}](D) = \begin{cases} 0, & D \geq 4.6 \\ \frac{1}{2} \ln \frac{4.6}{D}, & D \leq 4.6. \end{cases} \quad (41)$$

To determine $R[\text{Q-BOTH-W-NONE}](D)$ we must set up a constrained optimization as we did for the binary-Hamming scenario in Appendix B. This optimization results in a "water-pouring" bit allocation, which uses more bits to quantize the source when $q = 10$ than when $q = 1$. Specifically, the optimal test-channel is a Gaussian distribution where both the mean and the variance depend on $q$ and thus $\hat{x}$ has a Gaussian mixture distribution. Going through the details of the constrained optimization yields

$$R[\text{Q-BOTH-W-NONE}](D) = \begin{cases} 0, & 4.6 \leq D \\ \frac{0.4}{2} \ln \frac{4}{(D-.6)}, & D^* \leq D \leq 4.6 \\ \frac{0.4}{2} \ln \frac{10}{D} + \frac{0.6}{2} \ln \frac{1}{D}, & D \leq D^* \end{cases} \quad (42)$$

for some appropriate threshold $D^*$. Evaluating (34) for this case indicates that the rate-gap between (41) and (42) goes to $0.5 \cdot (\ln 4.6 - 0.4 \ln 10) \approx 0.3$ nats $\approx 0.43$ bits.

Computing $R[\text{Q-ENC-W-NONE}](D)$ analytically seems difficult. Thus, when distortion side information is only available at the encoder we obtain a numerical upper bound on the rate by using the same codebook distribution as when $q$ is known at both encoder and decoder. This yields a rate penalty of $I(\hat{x}; q)$.[12] We can obtain a simple analytic bound from Theorem 9. Specifically, evaluating (36) yields that the rate penalty is at most $(1/2) \cdot \min[1, D]$.

In Fig. 10 we evaluate these rate-distortion trade-offs. We see that at zero rate, the rate-distortion functions for the case of no side information, encoder-only side information, and full side information have the same distortion since no bits are available for quantization. Furthermore, we see that the Fisher Information bound is loose at zero rate. As the rate increases, the system with full distortion side-information does best because it uses the few available bits to represent only the important source samples with $q = 10$. The decoder reconstructs these source samples from the compressed data and reconstructs the less important samples to zero (the mean of $x$). In this regime, the system with distortion side information at the encoder also more accurately quantizes the important source samples. But since the decoder does not know $q$, it does not know which samples of $\hat{x}$ to reconstruct to zero. Thus the system with $q$ available at the encoder performs worse than the one with $q$ at both encoder and decoder but better than the system without side information. As the rate increases further, both systems with distortion side information quantize source samples with both $q = 1$ and $q = 10$. Thus the codebook distribution for $\hat{x}$ goes from a Gaussian mixture to become more and more Gaussian and the rate-loss for the system with only encoder side information goes to zero. Finally, we note that even at the modest distortion of $-5$ dB, the asymptotic effects promised by our theorems have already taken effect.

---
[12]Actually, since the rate distortion function is convex, we take the lower convex envelope of the curve resulting from the optimal test-channel distribution.

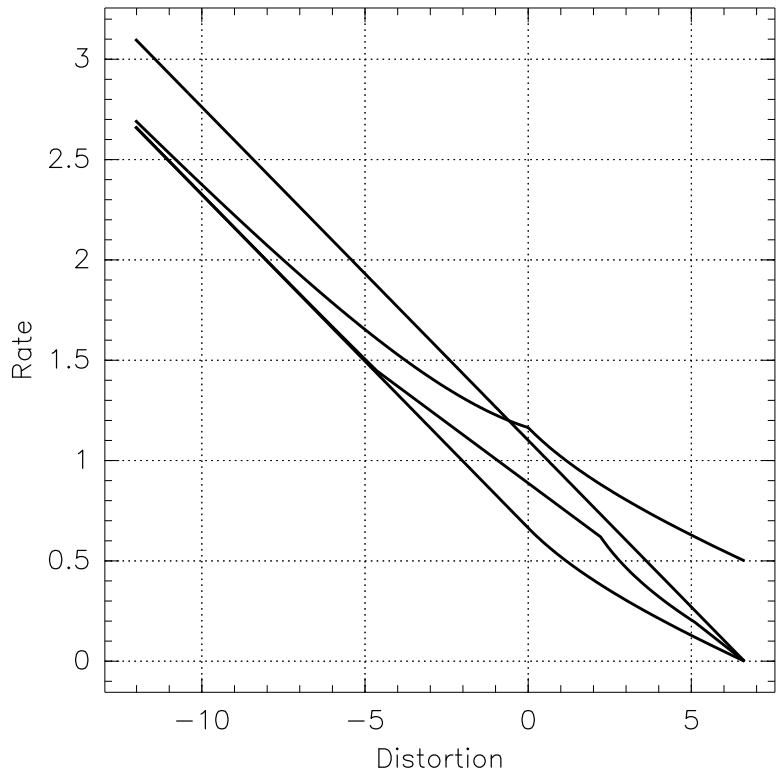

Fig. 10. Rate-distortion curves for quantizing a Gaussian source $x$ with distortion $q(x - \hat{x})^2$ where the side information $q$ is 1 with probability 0.6 or 10 with probability 0.4. From bottom to top on the right the curves correspond to the rate required when both encoder and decoder know $q$, a numerically computed upper bound to the rate when only the encoder knows $q$, the rate when neither encoder nor decoder know $q$, and the Fisher Information upper bound from Theorem 9 for when only the encoder knows $q$.

## VII. Discussion

In this paper, we introduced the notion of distortion side information, which does not directly depend on the source but instead affects the distortion measure. Furthermore, we showed that if general side information can be decomposed into such a distortion dependent component and a signal dependent component then under certain conditions the former is required only at the encoder and the latter is required only at the decoder. In this section, we discuss some implications and applications of these ideas.

### A. Applications to Sensors and Sensor Networks

There is a growing interest in sensor networks where multiple nodes with potentially correlated observations cooperate to sense the environment. A variety of researchers have already demonstrated the advantages of distributed source coding in efficiently using signal side information in such networks [27], [28], [29], [30], [31]. We believe, however, that the possible applications of distributed coding with distortion side information are equally compelling since many sensors naturally receive distortion side information in addition to the observed signal.

For example, sensors which perform simple filtering or averaging of the incoming signal can use the variance as an indicator of reliability. Similarly, sensors can observe the background level of light, sound, *etc.*, to obtain an estimate of the noise-floor in the absence of a signal of interest. Also, many systems are designed to record data at a roughly constant average amplitude. This is often accomplished by using automatic gain control (AGC) to compensate for attenuation due to distance, weather, or obstacles. Since thermal noise in the sensor front-end is usually independent of such attenuations, the AGC level can be used as a simple indicator of the signal-to-noise ratio of an observed signal.

### B. Richer Distortion Models and Perceptual Coding

The notion of distortion side information may be particularly useful in the design of perceptual coders. Specifically, it is well-known that mean-square error is at best a poor representative of how humans experience distortion. Hence perceptual coding systems use features of the human visual system (HVS) or human auditory system (HAS) to achieve low subjective distortion even when the mean-square distortion is quite large. Unfortunately, creating such a perceptual coder often requires the designer to be an expert both in human physiology, as well as quantizer design. This difficulty may be one of the reasons why information theory has sometimes had less impact on compression standards than on communication standards.

Using the abstraction of distortion side information to represent such perceptual effects, however, may help overcome this barrier. For example, physiological experts could focus on producing a distortion model that incorporates perceptual effects by determining a value $q_i$ that scales the quadratic distortion between the $i$th source sample and its reconstruction. Quantization experts could focus on designing compression systems operating with distortion side information. Then various perceptual models could be quickly and easily combined with quantizer designs to find the best combination. This modular design would allow a good quantization system to be used in a variety of application domains simply by changing the model for the distortion side information. Similarly, it would allow a finely tuned perceptual model to be used in many types of quantizer designs.

The theoretical justification for using distortion side information to design modular quantizers is Theorem 3. This theorem can be interpreted as saying that even if the perceptual model that produces the distortion weighting $q_i$ is a complicated function of the source, the decoder does not need to know how the perceptual model works. Instead, it is sufficient for the decoder simply to know the statistics of $q$ provided that $q$ is available to the encoder. Corollary 1 strengthens this conclusion since even when the source and distortion side information are statistically dependent, the gap between the modular, encoder-only side information architecture and a system with full side information is the mutual information between the source and the side information. No system (including non-modular systems where the decoder attempts to account for dependence between the source and distortion measure) could expect to do better since this rate gap essentially corresponds to how many bits about the source the side information implicitly conveys. Since no explicit side information is available at the decoder, no scheme can recover this rate gap.

Finally, it may be somewhat premature to advocate a particular design for perceptual coders based on primarily on rate-distortion results. At a minimum, however, we point out that if a perceptual distortion side information model can be constructed it can at least be used to find a bound on the minimum possible bit rate at a given distortion. Having such a performance benchmark to strive for can be a powerful force in inspiring system designer to search for new innovations.

### C. Decomposing Side Information Into **q** and **w**

Our problem model of Section III requires that the side information can be decomposed into distortion side information and signal side information. In many scenarios, this may be a natural view of the compression task. In problems where this is not the case, however, we briefly discuss how general side information may be decomposed into the pair $(\mathbf{q}, \mathbf{w})$ required by our theorems.

First, notice that the formulation in (3) is completely general in the sense that any side information $z$ taking values in the set $\mathcal{Z}$ can be trivially decomposed into $(q, w)$ by setting $(q, w) = (z, z)$ with $(\mathcal{Q}, \mathcal{W}) = (\mathcal{Z}, \mathcal{Z})$. Of course, this makes most of our results uninteresting. One systematic procedure for potentially improving this decomposition is as follows. First, replace each pair of values $(q_0, q_1)$ where $d(\cdot, \cdot; q_0) = d(\cdot, \cdot; q_1)$ with a new value $q'$ and adjust $\mathcal{Q}$ accordingly. Next, replace each pair of values $(w_0, w_1)$ where $p_{x|w}(\cdot|w_0) = p_{x|w}(\cdot|w_1)$ and $p_{q|w}(\cdot|w_0) = p_{q|w}(\cdot|w_1)$ with a new value $w'$ and adjust $\mathcal{W}$ accordingly. If the resulting $\mathcal{Q}$ and $\mathcal{W}$ are smaller than $\mathcal{Z}$, then our results become non-trivial.

As with the problem of determining a minimal sufficient statistics, there are many potential decompositions and different notions of good decompositions may be appropriate in different applications. For example, minimizing the cardinality of $\mathcal{Q}$, $\mathcal{W}$, or both, might be useful in simplifying quantizer design or related tasks. Alternatively, minimizing $I(\mathsf{q}; \mathsf{w})$, $H(\mathsf{q})$, $H(\mathsf{w})$, or similar information measures may be useful if **q**/**w** must be communicated to the encoder/decoder by a third party.

## VIII. Concluding Remarks

Our analysis indicates that side information that affects the distortion measure can provide significant benefits in source coding. Perhaps, our most surprising result is that in a number of cases, (*e.g.*, sources uniformly distributed over a group, or in the high-resolution limit) side information at the encoder is just as good as side information known at both encoder and decoder. Furthermore, this "separation theorem" can be composed with the previously known result that having signal side information at the decoder is often as good as having it both encoder and decoder (*e.g.*, in the high-resolution limit). Our main results regarding when knowing a given type of side information at one place is as good as knowing it at another place are summarized in Fig. 9. Also, we computed the rate-loss for lacking a particular type of side information in a specific place. These penalty theorems show that lacking the proper side information can produce arbitrarily large degradations in performance. Taken together, we believe these results suggest that distortion side information is a useful source coding paradigm.

## Appendix

### A. Group Difference Distortion Measures Proof

*Proof of Theorem 1:* Assume that $p^*_{\hat{x}|x,q}(\hat{x}|x, q)$ is an optimal test-channel distribution with the conditional $p^*_{\hat{x}|q}(\hat{x}|q)$. By symmetry, for any $t \in \mathcal{X}$, the shifted distribution

$$p^t_{\hat{x}|x,q}(\hat{x}|x, q) \stackrel{\Delta}{=} p^*_{\hat{x}|x,q}(\hat{x} \oplus t|x \oplus t, q) \tag{43}$$

must also be an optimal test-channel. Since mutual information is convex in the test-channel distribution, we obtain an optimal test-channel distribution $p^{**}$ by averaging $t$ over $\mathcal{X}$ via the uniform measure $d_{\mathcal{X}}(t)$:

$$p^{**}_{\hat{x}|x,q}(\hat{x}|x,q) \triangleq \int_{\mathcal{X}} p^{t}_{\hat{x}|x,q}(\hat{x}|x,q) d_{\mathcal{X}}(t). \tag{44}$$

To prove that the resulting distribution for $\hat{x}$ given $q$ is uniform for all $q$ (and hence independent of $q$), we will show that $p^{**}_{\hat{x}|q}(\hat{x}|q) = p^{**}_{\hat{x}|q}(\hat{x} \oplus r|q)$ for any $r \in \mathcal{X}$:

$$p^{**}_{\hat{x}|q}(\hat{x}|q) = \int_{\mathcal{X}} p^{**}_{\hat{x}|x,q}(\hat{x}|x,q) d_{\mathcal{X}}(x) \tag{45}$$

$$= \int_{\mathcal{X}} \int_{\mathcal{X}} p^{t}_{\hat{x}|x,q}(\hat{x}|x,q) d_{\mathcal{X}}(t) d_{\mathcal{X}}(x) \tag{46}$$

$$= \int_{\mathcal{X}} \int_{\mathcal{X}} p^{*}_{\hat{x}|x,q}(\hat{x} \oplus t|x \oplus t, q) d_{\mathcal{X}}(t) d_{\mathcal{X}}(x) \tag{47}$$

$$= \int_{\mathcal{X}} \int_{\mathcal{X}} p^{*}_{\hat{x}|x,q}(\hat{x} \oplus r \oplus t|x \oplus r \oplus t, q) d_{\mathcal{X}}(r \oplus t) d_{\mathcal{X}}(x) \tag{48}$$

$$= \int_{\mathcal{X}} \int_{\mathcal{X}} p^{*}_{\hat{x}|x,q}(\hat{x} \oplus r \oplus t|x \oplus r \oplus t, q) d_{\mathcal{X}}(t) d_{\mathcal{X}}(x \oplus r) \tag{49}$$

$$= \int_{\mathcal{X}} \int_{\mathcal{X}} p^{*}_{\hat{x}|x,q}(\hat{x} \oplus r \oplus t|x \oplus t, q) d_{\mathcal{X}}(t) d_{\mathcal{X}}(x) \tag{50}$$

$$= p^{**}_{\hat{x}|q}(\hat{x} \oplus r|q). \tag{51}$$

Equation (45) follows from Bayes' law and the fact that $d_{\mathcal{X}}$ is the uniform measure on $\mathcal{X}$. The next two lines follow from the definition of $p^{**}$ and $p^t$ respectively. To obtain (48), we make the change of variable $t \to r \oplus t$, and then apply the fact that the uniform measure is shift invariant to obtain (49). Similarly, we make the change of variable $x \oplus r \to x$ to obtain (50). The last line follows from the definition in (44).

Note that this argument applies regardless of whether the side information is available at the encoder, decoder, both, or neither. ∎

### B. Binary-Hamming Rate-Distortion Derivations

In this section we derive the rate-distortion functions for a binary source with Hamming distortion.

*1) With Encoder Side Information:* The rate-distortion function for source independent side information available only at the encoder is the same as with the side information available at both encoder and decoder. Hence, we compute $R[\text{Q-ENC}](D)$ and $R[\text{Q-BOTH}](D)$ by considering the latter case and noting that optimal encoding corresponds to simultaneous description of independent random variables [32,

Section 13.3.3]. Specifically, the source samples for each value of $q$ can be quantized separately using the distribution

$$p_{\hat{x}|x,q}(\hat{x}|x,q) = \begin{cases} 1 - p_q, & \hat{x} = x \\ p_q, & \hat{x} = 1 - x. \end{cases} \tag{52}$$

The cross-over probabilities $p_q$ correspond to the bit allocations for each value of the side information and are obtained by solving a constrained optimization problem:

$$R[\text{Q-BOTH}](D) = \min_{E[d(x,\hat{x};q)=D]} \sum_{i=1}^{N} E[1 - H_b(p_q)] \tag{53}$$

where $H_b(\cdot)$ is the binary entropy function.

Using Lagrange multipliers, we construct the functional

$$J(D) = \sum_{i=1}^{N} p_q(i) \cdot [1 + p_i \log p_i + (1 - p_i) \log(1 - p_i)] + \lambda \sum_{i=1}^{N} p_q(i) \cdot [\alpha_i + p_i \beta_i].$$

Differentiating with respect to $p_i$ and setting equal to 0, yields

$$\frac{\partial J}{\partial p_i} = p_q(i) \log \frac{p_i}{1 - p_i} + \lambda p_q(i) \beta_i = 0 \tag{54}$$

$$\log \frac{p_i}{1 - p_i} = -\lambda s_i \tag{55}$$

$$p_i = \frac{2^{-\lambda \beta_i}}{1 + 2^{-\lambda \beta_i}}. \tag{56}$$

Thus we obtain the rate-distortion functions

$$R[\text{Q-ENC}](D) = R[\text{Q-BOTH}](D) = 1 - \sum_{i=1}^{N} p_q(i) \cdot H_b\left(\frac{2^{-\lambda \beta_i}}{1 + 2^{-\lambda \beta_i}}\right) \tag{57a}$$

where $\lambda$ is chosen to satisfy

$$\sum_{i=1}^{N} p_q(i) \left[\alpha_i + \beta_i \cdot \frac{2^{-\lambda \beta_i}}{1 + 2^{-\lambda \beta_i}}\right] = D. \tag{57b}$$

*2) Without Encoder Side Information:* When no encoder side information is available, decoder side information is useless. Hence, the problem is equivalent to quantizing a symmetric binary source with distortion measure

$$d(x, \hat{x}) = E[\alpha_q + \beta_q \cdot d_H(x, \hat{x})] = E[\alpha_q] + E[\beta_q] \cdot d_H(x, \hat{x}). \tag{58}$$

The rate-distortion function is obtained by scaling and translating the rate-distortion function for the classical binary-Hamming case:

$$R[\text{Q-NONE}](D) = 1 - H_b\left(\frac{D - E[\alpha_q]}{E[\beta_q]}\right) \tag{59}$$

October 20, 2018 DRAFT

## C. High-Resolution Proofs

*Proof of Theorem 3:* To obtain $R[\text{Q-ENC-W-DEC}](D)$ we apply the Wyner-Ziv rate-distortion formula[13] to the "super-source" $\mathbf{x}' = (\mathbf{x}, \mathbf{q})$ yielding

$$R[\text{Q-ENC-W-DEC}](D) = \inf_{p_{\hat{x}|x,q}(\hat{x}|x,q)} I(\hat{x}, q; x|w) \tag{60}$$

where the optimization is subject to the constraint that $E[d(x, v(\hat{x}, w); q)] \leq D$ for some reconstruction function $v(\cdot, \cdot)$. To obtain $R[\text{Q-BOTH-W-BOTH}](D)$ we specialize the well-known conditional rate-distortion function to our notation yielding

$$R[\text{Q-BOTH-W-BOTH}](D) = \inf_{p_{\hat{x}|x,q,w}(\hat{x}|x,q,w)} I(\hat{x}; x|w, q) \tag{61}$$

where the optimization is subject to the constraint that $E[d(x, \hat{x}; q)] \leq D$.

Let us define $\hat{x}^*$ as the distribution that optimizes (60). Similarly, define $\hat{x}_w^*$ as the distribution that optimizes (61). Finally, define $z$ given $q = q$ to be a random variable with a conditional distribution that maximizes $h(z|q = q)$ subject to the constraint that

$$E[d(x, x + z; q)|q = q] \leq E[d(x, \hat{x}_w^*; q)|q = q]. \tag{62}$$

Then we have the following chain of inequalities:

$$\Delta R(D) \triangleq R[\text{Q-ENC-W-DEC}](D) - R[\text{Q-BOTH-W-BOTH}](D) \tag{63}$$

$$= I(\hat{x}^*; q, x|w) - [h(x|q, w) - h(x|q, w, \hat{x}_w^*)] \tag{64}$$

$$= I(\hat{x}^*; q, x|w) - h(x|q, w) + h(x - \hat{x}_w^*|q, w, \hat{x}_w^*) \tag{65}$$

$$\leq I(\hat{x}^*; q, x|w) - h(x|q, w) + h(x - \hat{x}_w^*|q) \tag{66}$$

$$\leq I(\hat{x}^*; q, x|w) - h(x|q, w) + h(z|q) \tag{67}$$

$$\leq I(x + z; q, x|w) - h(x|q, w) + h(z|q) \tag{68}$$

$$= h(x + z|w) - h(x + z|w, q, x) - h(x|q, w) + h(z|q) \tag{69}$$

$$= h(x + z|w) - h(x|q, w) \tag{70}$$

$$= h(x + z|w) - h(x|w) \tag{71}$$

$$\lim_{D \to D_{\min}} \Delta R(D) = 0. \tag{72}$$

---

[13]Some readers may be more familiar with the Wyner-Ziv formula as a difference of mutual informations (*e.g.*, as in [6]), but the form in (60) is equally valid [4] and is sometimes more convenient.

Equation (67) follows from the definition of z to be entropy maximizing subject to a distortion constraint. Since z is independent of x and w, the choice $\hat{x} = x + z$ with $v(\hat{x}, w) = \hat{x}$ is an upper bound to (60) and yields (68). We obtain (71) by recalling that according to our problem model in (4), q and x are independent given w. Finally, we obtain (72) by using the "continuity of entropy" result from [23, Theorem 1].

Note that although the z in [23, Theorem 1] is an entropy maximizing distribution while our z is a mixture of entropy maximizing distributions, the special form of the density is not required for the continuity of entropy result in [23, Theorem 1]. To illustrate this, we show how to establish the continuity of entropy directly for any distortion measure where $D \to D_{\min} \Rightarrow \mathrm{Var}[z] \to 0$. One example of such a distortion measure is obtained if we choose $d(x, \hat{x}; q) = q \cdot |x - \hat{x}|^r$ with $r > 0$ and $\Pr[q = 0] = 0$. Denoting $\mathrm{Var}\,[z|w]$ as $\sigma^2_{z|w}$ and $\mathrm{Var}\,[x|w]$ as $\sigma^2_{x|w}$ and letting $\mathcal{N}(\alpha)$ represent a Gaussian random variable with variance $\alpha$ yields

$$\limsup_{D \to D_{\min}} h(x+z|w) - h(x|w) = \limsup_{\sigma^2 \to 0} h(x+z|w) - h(x|w) \tag{73}$$

$$= \limsup_{\sigma^2 \to 0} h(x+z|w) \pm h(\mathcal{N}(\sigma^2_{x|w} + \sigma^2_{z|w})|w)$$
$$\pm h(\mathcal{N}(\sigma^2_{x|w})|w) - h(x|w) \tag{74}$$

$$= \limsup_{\sigma^2 \to 0} D(p_{x|w}||\mathcal{N}(\sigma^2_{x|w})) - D(p_{x+z|w}||\mathcal{N}(\sigma^2_{x|w} + \sigma^2_{z|w}))$$
$$+ h(\mathcal{N}(\sigma^2_{x|w} + \sigma^2_{z|w})|w) - h(\mathcal{N}(\sigma^2_{x|w})|w) \tag{75}$$

$$\leq D(p_{x|w}||\mathcal{N}(\sigma^2_{x|w})) - D(p_{x|w}||\mathcal{N}(\sigma^2_{x|w}))$$
$$+ \limsup_{\sigma^2 \to 0}[h(\mathcal{N}(\sigma^2_{x|w} + \sigma^2_{z|w})|w) - h(\mathcal{N}(\sigma^2_{x|w})|w)] \tag{76}$$

$$= \limsup_{\sigma^2 \to 0} \int \left[h(\mathcal{N}(\sigma^2_{x|w} + \sigma^2_{z|w})|w=w) - h(\mathcal{N}(\sigma^2_{x|w})|w=w)\right] dp_w(w) \tag{77}$$

$$= \int \left[\limsup_{\sigma^2 \to 0} h(\mathcal{N}(\sigma^2_{x|w} + \sigma^2_{z|w})|w=w) - h(\mathcal{N}(\sigma^2_{x|w})|w=w)\right] dp_w(w) \tag{78}$$

$$= 0. \tag{79}$$

We obtain (75) since for any random variable v, the relative entropy from v to a Gaussian takes the special form $D(p_v||\mathcal{N}(\mathrm{Var}\,[v])) = h(\mathcal{N}(\mathrm{Var}\,[v])) - h(v)$ [32, Theorem 9.6.5]. To get (76) we use the fact that relative-entropy (and also conditional relative-entropy) is lower semi-continuous [33]. This could also be shown by applying Fatou's Lemma [34, p.78] to get that if the sequences $p_1(x), p_2(x), \ldots$ and

$q_1(x), q_2(x), \ldots$ converge to $p(x)$ and $q(x)$ then

$$\liminf \int p_i(x) \log[p_i(x)/q_i(x)] \geq \int p(x) \log[p(x)/q(x)].$$

Switching the $\limsup$ and integral in (78) is justified by Lebesgue's Dominated Convergence Theorem [34, p.78] since the integrand is bounded for all values of $w$. In general, this bound is obtained from combining the technical condition requiring $h(x|w = w)$ to be finite with the entropy maximizing distribution in (25) and the expected distortion constraint in (26) to bound $h(x + z|q = q)$. For scaled quadratic distortions, $h(x + z|q = q)$ can be bounded above by the entropy of a Gaussian with the appropriate variance. To obtain (79) we first note that $\text{Var}[z] \to 0$ implies $\text{Var}[z|w = w] \to 0$ except possibly for a set of $w$ having measure zero. This set of measure zero can be ignored because the integrand is finite for all $w$. Finally, for the set of $w$ where $\text{Var}[z|w = w] \to 0$, the technical requirement that the entropy maximizing distribution in (25) is continuous shows that the entropy difference (79) goes to zero in the limit. ∎

*Proof of Theorem 4:* When $* \in \{\text{ENC}, \text{BOTH}\}$ in (28), the encoder can simulate **w** by generating it from $(\mathbf{x}, \mathbf{q})$. When $* \in \{\text{DEC}, \text{NONE}\}$, the encoder can still simulate **w** correctly provided that **w** and **q** are independent. Thus being provided with **w** provides no advantage given the conditions of the theorem. ∎

*Proof of Theorem 5:* We begin by showing

$$R[\text{Q-DEC-W-DEC}](D) = R[\text{Q-NONE-W-DEC}](D). \tag{80}$$

When side information $(\mathbf{q}, \mathbf{w})$ is available only at the decoder, the optimal strategy is Wyner-Ziv encoding [4]. Let us compute the optimal reconstruction function $v(\cdot, \cdot, \cdot)$, which maps an auxiliary random variable $u$ and the side information $q$ and $w$ to a reconstruction of the source:

$$v(u, q, w) = \arg\min_{\hat{x}} E[d(\hat{x}, \mathsf{x}; \mathsf{q}) | \mathsf{q} = q, \mathsf{w} = w, \mathsf{u} = u] \tag{81}$$

$$= \arg\min_{\hat{x}} d_0(q) E[d_1(\hat{x}, \mathsf{x}) | \mathsf{q} = q, \mathsf{w} = w, \mathsf{u} = u] \tag{82}$$

$$= \arg\min_{\hat{x}} E[d_1(\hat{x}, \mathsf{x}) | \mathsf{q} = q, \mathsf{w} = w, \mathsf{u} = u] \tag{83}$$

$$= \arg\min_{\hat{x}} E[d_1(\hat{x}, \mathsf{x}) | \mathsf{w} = w, \mathsf{u} = u]. \tag{84}$$

We obtain (82) from the assumption that we have a separable distortion measure. To get (84) recall that by assumption $q$ is statistically independent of $x$ given $w$ and also $q$ is statistically independent of $u$ since $u$ is generated at the encoder from $x$. Thus neither the optimal reconstruction function $v(\cdot, \cdot, \cdot)$ nor the auxiliary random variable $u$ depend on $q$. This establishes (80).

To show that
$$R[\text{Q-DEC-W-NONE}](D) = R[\text{Q-NONE-W-NONE}](D) \tag{85}$$
we need **w** and **q** to be independent. When this is true, **w** does not affect anything and the problem is equivalent to when **w** = 0 and is available at the decoder. From (80) we see that providing **w** = 0 at the decoder does not help and thus we establish (85). Note that this argument fails when **w** and **q** are not independent since in that case Wyner-Ziv based on **q** could be performed and there would be no **w** at the decoder to enable the argument in (81)–(84).

To show that
$$R[\text{Q-DEC-W-BOTH}](D) = R[\text{Q-NONE-W-BOTH}](D) \tag{86}$$
we note that in this scenario the encoder and decoder can design a different source coding system for each value of $w$. The subsystem for a fixed value $w^*$ corresponds to source coding with distortion side information at the decoder. Specifically, the source will have distribution $p_{\mathsf{x}|\mathsf{w}}(x|w^*)$, and the distortion side information will have distribution $p_{\mathsf{q}|\mathsf{w}}(q|w^*)$. Thus the performance of each subsystem is given by $R[\text{Q-DEC-W-NONE}](D)$, which we already showed is the same as $R[\text{Q-NONE-W-NONE}](D)$. This establishes (86).

Finally, to show that
$$R[\text{Q-DEC-W-ENC}](D) = R[\text{Q-NONE-W-ENC}](D) \tag{87}$$
we require the assumption that **q** and **w** are independent. This assumption implies
$$R[\text{Q-DEC-W-ENC}](D) = R[\text{Q-DEC-W-NONE}](D) \tag{88}$$
since an encoder without **w** could always generated a simulated **w** with the correct distribution relative to the other variables. The same argument implies
$$R[\text{Q-NONE-W-ENC}](D) = R[\text{Q-NONE-W-NONE}](D). \tag{89}$$
Combining (88), (89), and (85) yields (87). ■

*Proof of Theorem 6:* First we establish the four rate-distortion function equalities implied by (30a). Using Theorem 3 we have
$$\lim_{D \to D_{\min}} R[\text{Q-ENC-W-DEC}](D) - R[\text{Q-BOTH-W-DEC}](D) \leq \tag{90}$$
$$\lim_{D \to D_{\min}} R[\text{Q-ENC-W-DEC}](D) - R[\text{Q-BOTH-W-BOTH}](D) \tag{91}$$
$$= 0. \tag{92}$$

Similarly,

$$\lim_{D \to D_{\min}} R[\text{Q-ENC-W-BOTH}](D) - R[\text{Q-BOTH-W-BOTH}](D) \leq \qquad (93)$$

$$\lim_{D \to D_{\min}} R[\text{Q-ENC-W-DEC}](D) - R[\text{Q-BOTH-W-BOTH}](D) \qquad (94)$$

$$= 0. \qquad (95)$$

To show that

$$\lim_{D \to D_{\min}} R[\text{Q-ENC-W-NONE}](D) - R[\text{Q-BOTH-W-NONE}](D) = 0 \qquad (96)$$

we need **q** and **w** to be independent. When this is true, **w** does not affect anything and the problem is equivalent to when **w** = 0 and is available at the decoder and (90)–(92) establishes (96). Without independence this argument fails because we can no longer invoke Theorem 3 since there will be no **w** to make **x** and **q** conditionally independent in (71).

To finish establishing (30a) we again require **q** and **w** to be independent to obtain

$$\lim_{D \to D_{\min}} R[\text{Q-ENC-W-ENC}](D) - R[\text{Q-BOTH-W-ENC}](D) \leq \qquad (97)$$

$$\lim_{D \to D_{\min}} R[\text{Q-ENC-W-NONE}](D) - R[\text{Q-BOTH-W-ENC}](D) = \qquad (98)$$

$$\lim_{D \to D_{\min}} R[\text{Q-ENC-W-NONE}](D) - R[\text{Q-BOTH-W-NONE}](D) \qquad (99)$$

$$= 0 \qquad (100)$$

where (99) follows since the encoder can always simulate **w** from $(\mathbf{x}, \mathbf{q})$ and (100) follows from (96).

Next, we establish the four rate-distortion function equalities implied by (30b). Using Theorem 3 we have

$$\lim_{D \to D_{\min}} R[\text{Q-ENC-W-DEC}](D) - R[\text{Q-ENC-W-BOTH}](D) \leq \qquad (101)$$

$$\lim_{D \to D_{\min}} R[\text{Q-ENC-W-DEC}](D) - R[\text{Q-BOTH-W-BOTH}](D) \qquad (102)$$

$$= 0. \qquad (103)$$

Similarly,

$$\lim_{D \to D_{\min}} R[\text{Q-BOTH-W-DEC}](D) - R[\text{Q-BOTH-W-BOTH}](D) \leq \qquad (104)$$

$$\lim_{D \to D_{\min}} R[\text{Q-ENC-W-DEC}](D) - R[\text{Q-BOTH-W-BOTH}](D) \qquad (105)$$

$$= 0. \qquad (106)$$

To show that

$$\lim_{D \to D_{\min}} R[\text{Q-NONE-W-DEC}](D) - R[\text{Q-NONE-W-BOTH}](D) = 0 \qquad (107)$$

we need **q** and **w** to be independent and we need the distortion measure to be of the form $d(\hat{x}, x; q) = d_0(q) \cdot d_1(x - \hat{x})$. When this is true the two rate-distortion functions in (107) are equivalent to the Wyner-Ziv rate-distortion function and the conditional rate-distortion function for the difference distortion measure $E[d_0(q)] \cdot d_1(x - \hat{x})$. Thus we can either apply the result from [8] showing these rate-distortion functions are equal in the high-resolution limit or simply specialize Theorem 3 to the case where **q** is a constant.

To complete the proof, we again require the assumptions that **q** and **w** are independent and that the distortion measure is of the form $d(x, \hat{x}; q) = d_0(q) \cdot d_0(q) \cdot d_1(x - \hat{x})$. We have

$$\lim_{D \to D_{\min}} R[\text{Q-DEC-W-DEC}](D) - R[\text{Q-DEC-W-BOTH}](D) \leq \qquad (108)$$

$$\lim_{D \to D_{\min}} R[\text{Q-NONE-W-DEC}](D) - R[\text{Q-DEC-W-BOTH}](D) = \qquad (109)$$

$$\lim_{D \to D_{\min}} R[\text{Q-NONE-W-DEC}](D) - R[\text{Q-NONE-W-BOTH}](D) \qquad (110)$$

$$= 0. \qquad (111)$$

where (110) follows from Theorem 5 and (111) follows from (107). ∎

*Proof of Theorem 7:* We note that according to Theorem 4 and Theorem 6 we can focus solely on the case

$$R[\text{Q-*-W-NONE}](D) - R[\text{Q-*-W-BOTH}](D). \qquad (112)$$

When * = NONE, the rate difference in (112) is the difference between the classical rate-distortion function and the conditional rate-distortion function in the high-resolution limit. Thus the Shannon Lower Bound [23] (and its conditional version) imply that

$$\lim_{D \to D_{\min}} R[\text{Q-NONE-W-NONE}](D) - R[\text{Q-NONE-W-BOTH}](D) = h(x) - h(x|w). \qquad (113)$$

Similarly, when * = DEC an identical argument can be combined with Theorem 5.

When * = BOTH, the encoder and decoder can design a separate compression sub-system for each value of *q*. The rate-loss for each sub-system is then $I(x; w | q = q)$ according to high-resolution Wyner-Ziv theory [8]. Averaging over all values of *q* yields a total rate-loss of $I(x; w | q)$.

Next we consider the case when * = ENC and the rate-loss penalty is

$$\lim_{D \to D_{\min}} R[\text{Q-ENC-W-NONE}](D) - R[\text{Q-ENC-W-BOTH}](D)$$

$$= \lim_{D \to D_{\min}} R[\text{Q-ENC-W-NONE}](D) - R[\text{Q-BOTH-W-BOTH}](D) \qquad (114)$$

where the equality follows from Theorem 6.

Using arguments similar to [23] and the proof of Theorem 3, we can obtain a Shannon Lower Bound for $R[\text{Q-ENC-W-NONE}](D)$, which is of the form

$$R[\text{Q-ENC-W-NONE}](D) \geq h(\mathsf{x}) - h(\mathsf{z}_D) \tag{115}$$

where $\mathsf{z}_D$ is an entropy maximizing random variable subject to the constraint that $E[d_0(\mathsf{q}) \cdot d_1(\mathsf{z}_D)] \leq D$. Again using argument similar to the proof of Theorem 3, we have that

$$\lim_{D \to D_{\min}} R[\text{Q-BOTH-W-BOTH}](D) \leq h(\mathsf{x}|\mathsf{w}) - h(\mathsf{z}_D). \tag{116}$$

Combining (115) and (116) shows that the asymptotic difference in (114) is at least $I(\mathsf{x};\mathsf{w})$.

Next, we obtain the Shannon Lower Bound

$$R[\text{Q-BOTH-W-BOTH}](D) \geq h(\mathsf{x}|\mathsf{w}) - h(\mathsf{z}_D) \tag{117}$$

by duplicating the arguments in the proof of Theorem 3 since this lower bound does not require $\mathsf{q}$ and $\mathsf{w}$ to be independent. Finally, we can obtain the upper bound

$$\lim_{D \to D_{\min}} R[\text{Q-ENC-W-NONE}](D) \leq h(\mathsf{x}) - h(\mathsf{z}_D) \tag{118}$$

using an additive noise test channel combined with arguments following those in the proof of Theorem 3. Combining (117) and (118) shows that the asymptotic difference in (114) is at most $I(\mathsf{x};\mathsf{w})$.

∎

*Proof of Theorem 8:* To simplify the exposition, we first prove the theorem for the relatively simple case of a one-dimensional source ($k = 1$) with a quadratic distortion ($r = 2$). Then at the end of the proof, we describe how to extend it to general $k$ and $r$.

We begin with the case where * = NONE. Since Theorems 5 and 6 imply

$$R[\text{Q-NONE-W-NONE}](D) = R[\text{Q-DEC-W-NONE}](D) \tag{119a}$$

and

$$R[\text{Q-ENC-W-NONE}](D) \to R[\text{Q-BOTH-W-NONE}](D) , \tag{119b}$$

we focus on showing

$$\lim_{D \to D_{\min}} R[\text{Q-BOTH-W-NONE}](D) - R[\text{Q-NONE-W-NONE}](D) = \frac{1}{2} E\left[\ln \frac{E[\mathsf{q}]}{\mathsf{q}}\right]. \tag{120}$$

Computing $R[\text{Q-BOTH-W-NONE}](D)$ is equivalent to finding the rate-distortion function for optimally encoding independent random variables and yields the familiar "water-pouring" rate and distortion allocation [32, Section 13.3.3]. For each $q$, we quantize the corresponding source samples with distortion $D_q = E[(x - \hat{x})^2]$ (or $E[||\mathbf{x} - \hat{\mathbf{x}}||^r]$ in the more general case) and rate $R_q(D_q)$. The overall rate and distortion then become $E[R_q(D_q)]$ and $E[q \cdot D_q]$.

Thus to find the rate and distortion allocation we set up a constrained optimization problem using Lagrange multipliers to obtain the functional

$$J(D) = E[R_q(D_q)] + \lambda(D - E[q \cdot D_q]), \tag{121}$$

differentiate with respect to $D_q$, set equal to zero and solve for each $D_q$. In the high-resolution limit, various researchers have shown

$$R_q(D_q) \to h(x) - \frac{1}{2} \log D_q. \tag{122}$$

(*e.g.*, see [23] and references therein). Therefore, it is straightforward to show this process yields the condition $D_q = 1/(2\lambda q)$ with $2\lambda = 1/D$ implying

$$\lim_{D \to 0} R[\text{Q-BOTH-W-NONE}](D) \to h(x) - \frac{1}{2} \log D + \frac{1}{2} E[\log q]. \tag{123}$$

To compute $R[\text{Q-NONE-W-NONE}](D)$, we note that since neither encoder nor decoder knows $q$ the optimal strategy is to simply quantize the source according to the distortion $d(q, x; \hat{x}) = E[q] \cdot (x - \hat{x})^2$ to obtain

$$\lim_{D \to 0} R[\text{Q-NONE-W-NONE}](D) \to h(x) - \frac{1}{2} \log D + \frac{1}{2} \log E[q]. \tag{124}$$

Comparing (123) with (124) establishes (120).

By applying Theorem 4 we see that the case where $* =$ ENC is the same as $* =$ NONE.

Next we consider the case where $* =$ BOTH in (34). In this case, the encoder and decoder can design a separate compression sub-system for each value of $w$ and the performance for each sub-system is obtained from the case with no signal side information. Specifically, the rate-loss for each sub-system is

$$\frac{1}{2} E \left[ \ln \frac{E[q|w=w]}{q} \middle| w = w \right] \tag{125}$$

according to the previously derived results. Averaging (125) over $w$ then yields the rate-loss in (34).

Finally, we consider the case where $* =$ DEC in (34). Since Theorem 5 implies $R[\text{Q-DEC-W-DEC}](D) = R[\text{Q-NONE-W-DEC}](D)$ and Theorem 3 implies $R[\text{Q-ENC-W-DEC}](D) \to R[\text{Q-BOTH-W-BOTH}](D)$, it suffices to show that

$$\lim_{D \to D_{\min}} R[\text{Q-DEC-W-DEC}](D) - R[\text{Q-BOTH-W-BOTH}](D) = \frac{1}{2} E \left[ \log \frac{E[q]}{q} \right]. \tag{126}$$

We can compute $R[\text{Q-BOTH-W-BOTH}](D)$ by considering a separate coding system for each value of $w$. Specifically, conditioned on $w = w$, computing the rate-distortion trade-off is equivalent to finding $R[\text{Q-BOTH-W-NONE}](D)$ for a modified source $x'$ with distribution $p_{x'}(x') = p_{x|w}(x'|w)$. Thus we obtain

$$\lim_{D \to D_{\min}} R[\text{Q-BOTH-W-BOTH}](D) \to h(x|w) - \frac{1}{2} \log D + \frac{1}{2} \log E[q]. \tag{127}$$

Applying the standard techniques used throughout the paper, we can compute the Shannon Lower Bound

$$R[\text{Q-DEC-W-DEC}](D) \geq h(x|w) - \frac{1}{2} \log(D \cdot E[q]) \tag{128}$$

and show it is tight in the high-resolution limit. Comparing (127) and (128) establishes the desired result.

This establishes the theorem for $k = 1$ and $r = 2$. For general $k$ and $r$, the only change is that each component rate-distortion function $R_q(D_q)$ (122) becomes [23, page 2028]

$$R_q(D_q) \to h(x) - \frac{k}{r} \log D_q - \frac{k}{r} + \log \left[ \frac{r}{k V_k \Gamma(k/r)} \left( \frac{k}{r} \right)^{k/r} \right]. \tag{129}$$

and a similar change occurs for all the following rate-distortion expressions. Since we are mainly interested in the difference of rate-distortion functions, most of these extra terms cancel out and the only change is that factors of $1/2$ are replaced with factors of $k/r$. ∎

### D. Proofs for Non-Asymptotic Bounds

Before proceeding, we require the following lemma to upper and lower bound the entropy of an arbitrary random variable plus a Gaussian mixture.

**Lemma 2** *Let $x$ be an arbitrary random variable with finite variance $\sigma^2 < \infty$. Let $z$ be a zero-mean, unit-variance Gaussian independent of $x$ and let $v$ be a random variable independent of $x$ and $z$ with $0 < v_{\min} \leq v < v_{\max}$. Then*

$$h(x) + \frac{1}{2} \log(1 + v_{\min}) \leq h(x + z\sqrt{v}) \leq h(x) + \frac{1}{2} \log(1 + v_{\max} \cdot J(x)) \tag{130}$$

*with equality if and only if $v$ is a constant and $x$ is Gaussian.*

*Proof:* The concavity of differential entropy yields

$$h(x + z\sqrt{v_{\min}}) \leq h(x + z\sqrt{v}) \leq h(x + z\sqrt{v_{\max}}). \tag{131}$$

For the lower bound we have

$$h(x + z\sqrt{v_{\min}}) = \int_0^{v_{\min}} \frac{\partial}{\partial \tau} h(x + z\sqrt{\tau})d\tau + h(x) \quad (132)$$

$$= \int_0^{v_{\min}} \frac{1}{2} J(x + z\sqrt{\tau})d\tau + h(x) \quad (133)$$

$$\geq \frac{1}{2}\int_0^{v_{\min}} J(z\sqrt{\sigma^2 + \tau})d\tau + h(x) \quad (134)$$

$$= \frac{1}{2}\int_0^{v_{\min}} \frac{d\tau}{\sigma^2 + \tau} + h(x) \quad (135)$$

$$= \frac{1}{2}\log\left(1 + \frac{v_{\min}}{\sigma^2}\right) + h(x) \quad (136)$$

where (133) follows from de Bruijn's identity [32, Theorem 16.6.2], [35, Theorem 14], (134) follows from the fact that a Gaussian distribution minimizes Fisher Information subject to a variance constraint, and (135) follows since the Fisher Information for a Gaussian is the reciprocal of its variance.

Similarly, for the upper bound we have

$$h(x + z\sqrt{v_{\max}}) = \int_0^{v_{\max}} \frac{\partial}{\partial \tau} h(x + z\sqrt{\tau})d\tau + h(x) \quad (137)$$

$$= \int_0^{v_{\max}} \frac{1}{2} J(x + z\sqrt{\tau})d\tau + h(x) \quad (138)$$

$$\leq \frac{1}{2}\int_0^{v_{\max}} \frac{J(x)J(z\sqrt{\tau})}{J(x) + J(z\sqrt{\tau})}d\tau + h(x) \quad (139)$$

$$= \frac{1}{2}\int_0^{v_{\max}} \frac{J(x)d\tau}{\tau J(x) + 1} + h(x) \quad (140)$$

$$= \frac{1}{2}\log\left(1 + v_{\max}\cdot J(x)\right) + h(x) \quad (141)$$

where (138) again follows from de Bruijn's identity, (139) follows from the convolution inequality for Fisher Information [36], [32, p.497], and (140) follows since the Fisher Information for a Gaussian is the reciprocal of its variance.

Combining these upper and lower bounds yields the desired result. Finally, the inequalities used in (134) and (139) are both tight if and only if $x$ is Gaussian. ∎

As an aside we note that Lemma 2 can be used to bound the rate-distortion function of an arbitrary unit-variance source $x$ relative to quadratic distortion. Specifically using an additive Gaussian noise test-channel $\hat{x} = z + x$ and combining Lemma 2 to upper bound $h(x + z)$ with the Shannon Lower Bound [23] yields

$$h(x) - \frac{1}{2}\log 2\pi e D \leq R(D) \leq h(x) - \frac{1}{2}\log 2\pi e D + \frac{1}{2}\log[1 + DJ(x)]. \quad (142)$$

Evidently, the error in the Shannon Lower Bound is at most $\frac{1}{2}\log[1 + DJ(x)]$. Thus, since $J(x) \geq 1$ with equality only for a Gaussian, the sub-optimality of an additive Gaussian noise test-channel is at least $\frac{1}{2}\log[1 + D]$.

*Proof of Theorem 9:* Starting with the bound for the rate gap from (71), we have

$$R[\text{Q-ENC-W-DEC}](D) - R[\text{Q-BOTH-W-BOTH}](D) \leq h(x + z|w) - h(x|w) \tag{143}$$

$$= \int \left[h(x + z|w = w) - h(x|w = w)\right] p_w(w) dw \tag{144}$$

$$\leq \int \left\{ \frac{1}{2} \log \left( 1 + \min\left[1, \frac{D}{q_{\min}}\right] \cdot J(x|w = w) \right) \right\} p_w(w) dw \tag{145}$$

$$\leq \int \left\{ \frac{J(x|w = w)}{2} \cdot \min\left[1, \frac{D}{q_{\min}}\right] \right\} p_w(w) dw \tag{146}$$

$$= \frac{J(x|w)}{2} \cdot \min\left[1, \frac{D}{q_{\min}}\right]. \tag{147}$$

To obtain (145) we note that $z$ is a Gaussian mixture and apply Lemma 2. This follows since, conditioned on $q = q$, $z$ is a Gaussian with variance $E[d(x, \hat{x}_w^*; q)]$ where $\hat{x}_w^*$ was defined in the proof of Theorem 3 to be the optimal distribution when both encoder and decoder know the side information. By considering the optimal "water-pouring" distortion allocation for the optimal test-channel distribution $\hat{x}_w^*$, it can be demonstrated that if the distortion is $D$, then $E[d(x, \hat{x}_w^*; q)]$ is at most $\min[1, D/q]$ for each $q$. ∎

To develop a similar bound for other distortion measures essentially all we need is an upper bound for the derivative of $h(x + \sqrt{\tau}z)$ with respect to $\tau$. Since entropy is concave, if we can compute this derivative for $\tau = 0$ then it will be an upper bound for the derivative at all $\tau$.

To obtain the desired derivative at $\tau = 0$, we can write

$$h(x + \sqrt{\tau}z) = I(x + \sqrt{\tau}z; \sqrt{\tau}z) - h(x). \tag{148}$$

The results of Prelov and van der Meulen [37] imply that under certain regularity conditions

$$\frac{\partial}{\partial \tau} \lim_{\tau \to 0^+} I(x + \sqrt{\tau}z; \sqrt{\tau}z) = J(x)/2 \; , \tag{149}$$

which provides the desired derivative. Similarly if we rewrite the mutual information in (148) as a relative entropy, then a Taylor series expansion of the relative entropy [38, 2.6] can be used to establish (149) provided certain derivatives of the probability distributions exist.

Next, we move to proving Theorem 10. An essential part of our proof is an alternative version of the Shannon Lower Bound, which we develop in the following lemma.

**Lemma 3 (Alternative Shannon Lower Bound)** *Consider a scaled quadratic distortion measure of the form $d(x, \hat{x}; q) = q \cdot (x - \hat{x})^2$ and let $\hat{x}^*_{q,w}$ denote an optimal test-channel distribution when $q$ and $w$ are known at both encoder and decoder. If we define $z$ to have the same distribution as $x - \hat{x}^*_{q,w}$ when conditioned on $q$ and furthermore require $z$ to satisfy the Markov condition $z \leftrightarrow q \leftrightarrow w, x$, then*

$$R[\text{Q-BOTH-W-BOTH}](D) \geq h(x|w) - h(z|q). \tag{150}$$

*Proof:*

$$R[\text{Q-BOTH-W-BOTH}](D) = I(\hat{x}^*_{q,w}; x|q, w) \tag{151}$$

$$= h(x|q, w) - h(x|q, w, \hat{x}^*_{q,w}) \tag{152}$$

$$= h(x|q, w) - h(x - \hat{x}^*_{q,w}|q, w, \hat{x}^*_{q,w}) \tag{153}$$

$$= h(x|q, w) - h(z|q, w, \hat{x}^*_{q,w}) \tag{154}$$

$$\geq h(x|q, w) - h(z|q, w). \tag{155}$$

∎

The key difference between Lemma 3 and the traditional Shannon Lower Bound (SLB) is in the choice of the distribution for $z$. The traditional SLB uses an entropy maximizing distribution for $z$, which has the advantage of being computable without knowing $\hat{x}^*_{q,w}$. The trouble with the entropy maximizing distribution is that it can have an unbounded variance for large distortions. As we show in the following lemma, however, the alternative SLB keeps the variance of $z$ bounded.

**Lemma 4** *There exists a choice for $z$ in Lemma 3 such that for all values of $w$,*

$$\text{Var}[z|w = w] \leq \text{Var}[x|w = w]. \tag{156}$$

*Proof:* Imagine that we choose some optimal test-channel distribution $\hat{x}^*_{q,w}$ such that the resulting $z$ does not satisfy (156) for some value of $w$. We will show that it is possible to construct an alternative optimal test-channel distribution $\hat{x}^{*\prime}_{q,w}$ where the resulting $z'$ does satisfy (156) for $w = w$.

Specifically, if (156) is not satisfied, then it must be that there exists a set $\mathcal{A}$ with

$$\text{Var}[z|q = q, w = w] = \text{Var}[x - \hat{x}^*_{q,w}|q = q, w = w] > \text{Var}[x|w = w], \forall (q, w) \in \mathcal{A}. \tag{157}$$

Define a new random variable $\hat{x}^{*\prime}_{q,w}$ such that $\hat{x}^{*\prime}_{q,w} = \hat{x}^*_{q,w}$ for all $(q, w) \notin \mathcal{A}$, but with $\hat{x}^{*\prime}_{q,w} = 0$ for all $(q, w) \in \mathcal{A}$. The distortion is lower for $\hat{x}^{*\prime}_{q,w}$ by construction. Furthermore, the date processing inequality implies that

$$I(\hat{x}^{*\prime}_{q,w}; x|w, q) \leq I(\hat{x}^*_{q,w}; x|w, q) \tag{158}$$

and so the rate is lower too. Thus if we define $z' = \hat{x}_{q,w}^{*\prime} - x$ analogously to how we defined $z$, then condition (156) is satisfied with $z$ replaced by $z'$. ∎

*Proof of Theorem 10:* Using the alternative SLB from Lemma 3 and the test-channel distribution $\hat{x} = x + z$ with $z$ chosen according to Lemma 3 we obtain

$$R[\text{Q-ENC-W-DEC}](D) - R[\text{Q-BOTH-W-BOTH}](D) \leq I(x+z; q, x|w) - [h(x|w, q) - h(z|w, q)] \tag{159}$$

$$= h(x+z|w) - h(x+z|q, x, w) - h(x|w, q) + h(z|w, q) \tag{160}$$

$$= h(x+z|w) - h(z|q, x, w) - h(x|w, q) + h(z|w, q) \tag{161}$$

$$= h(x+z|w) - h(z|q) - h(x|w, q) + h(z|q) \tag{162}$$

$$= h(x+z|w) - h(x|w) \tag{163}$$

$$= D(p_{x|w} \| \mathcal{N}(\sigma_{x|w}^2)) - D(p_{x+z|w} \| \mathcal{N}(\sigma_{x|w}^2 + \sigma_{z|w}^2))$$
$$+ h(\mathcal{N}(\sigma_{x|w}^2 + \sigma_{z|w}^2)|w) - h(\mathcal{N}(\sigma_{x|w}^2)|w) \tag{164}$$

$$\leq D(p_{x|w} \| \mathcal{N}(\sigma_{x|w}^2)) + h(\mathcal{N}(\sigma_{x|w}^2 + \sigma_{z|w}^2)|w) - h(\mathcal{N}(\sigma_{x|w}^2)|w) \tag{165}$$

$$= D(p_{x|w} \| \mathcal{N}(\sigma_{x|w}^2))$$
$$+ \int \left[ h(\mathcal{N}(\sigma_{x|w}^2 + \sigma_{z|w}^2)|w = w) - h(\mathcal{N}(\sigma_{x|w}^2)|w = w) \right] p_w(w) dw \tag{166}$$

$$\leq D(p_{x|w} \| \mathcal{N}(\sigma_{x|w}^2)) + \int \left[ \frac{1}{2} \log\left(1 + \frac{\sigma_{z|w}^2}{\sigma_{x|w}^2}\right) \right] p_w(w) dw \tag{167}$$

$$\leq D(p_{x|w} \| \mathcal{N}(\sigma_{x|w}^2)) + \int \left[ \frac{1}{2} \log\left(1 + \frac{\sigma_{\max}^2}{\sigma_{x|w}^2}\right) \right] p_w(w) dw \tag{168}$$

$$\leq D(p_{x|w} \| \mathcal{N}(\sigma_{x|w}^2)) + \int \left[ \frac{1}{2} \log\left(1 + \frac{\sigma_{\max}^2}{\sigma_{\min}^2}\right) \right] p_w(w) dw \tag{169}$$

$$= D(p_{x|w} \| \mathcal{N}(\sigma_{x|w}^2)) + \frac{1}{2} \log\left(1 + \frac{\sigma_{\max}^2}{\sigma_{\min}^2}\right). \tag{170}$$

To obtain (163)–(167) we use the same arguments as in (73)–(79) plus the additional observation that relative entropy is positive and can be dropped in obtaining (165). Next, we apply Lemma 4 to keep the variance of the test-channel noise to be at most $\sigma_{\max}^2$ to get (168). Finally, the assumption that $\sigma_{x|w}^2 \geq \sigma_{\min}^2$ yields (169). ∎

To develop a similar bound for other distortion measures, we would use an entropy maximizing distribution for the appropriate distortion measure in $D(p_{x|w} \| \cdot)$ and $D(p_{x+z} \| \cdot)$ above.